\documentclass[eqsecnum,aps,article]{revtex4}
\newcommand{\nwc}{\newcommand}
\nwc{\be}{\begin{equation}}
\nwc{\ee}{\end{equation}}
\nwc{\bea}{\begin{eqnarray}}
\nwc{\eea}{\end{eqnarray}}
\nwc{\cH}{{\cal H}}
\nwc{\cB}{{\cal B}}
\nwc{\cR}{{\cal R}}
\nwc{\cU}{{\cal U}}
\nwc{\Om}{\Omega}
\nwc{\cC}{{\cal C}}
\nwc{\cD}{{\cal D}}
\nwc{\rP}{{\rm P}}
\nwc{\chii}{{\underline {\overline \chi}}}

\begin{document}
\title{Geometric phase for mixed states: a differential geometric approach}
\author{S. Chaturvedi\footnote{email: scsp@uohyd.ernet.in}}
\address{Department of Physics, University of Hyderabad, Hyderabad 500046, 
India}
\author{E. Ercolessi\footnote{email: ercolessi@bo.infn.it}}
\address{Dipartimento di Fisica, Universita di Bologna, INFM and INFN, Via 
Irnerio 46, 40126 Bologna, Italy}
\author{G. Marmo\footnote{email: gimarmo@na.infn.it}}
\address{Dipartimento di Scienze Fisiche, Universita di Napoli Federico II and 
INFN, Via Cinzia, 80126 Napoli, Italy}
\author{G. Morandi\footnote{email: morandi@bo.infn.it}}
\address{Dipartimento di Fisica, Universita di Bologna, INFM and INFN, Viale 
Berti-Pichat 6/2, 40127 Bologna, Italy} 
\author{N. Mukunda\footnote{email: nmukunda@cts.iisc.ernet.in}}
\address{Centre for Theoretical Studies, Indian Institute of Science, 
Bangalore 560~012, India}
\author{R. Simon\footnote{email: simon@imsc.res.in}}
\address{The Institute of Mathematical Sciences, C. I. T. Campus, Tharamani, 
Chennai 600 113, India}
\begin{abstract}
A new definition and interpretation of geometric phase for mixed state cyclic
unitary evolution in quantum mechanics are presented. The pure state case is
formulated in a framework involving three selected Principal Fibre Bundles,
and the well known Kostant-Kirillov-Souriau symplectic structure on (co)
adjoint orbits associated with Lie groups. It is shown that this framework 
generalises in a natural and simple manner to the mixed state case. For 
simplicity, only the case of rank two mixed state density matrices is 
considered in detail. The extensions of the ideas of Null Phase Curves and 
Pancharatnam lifts from pure to mixed states are also presented.
\end{abstract}
\maketitle
\section{Introduction}
The theory of the geometric phase (GP) for pure state unitary
quantum evolution \cite{1} attained a definitive status in all 
essential aspects quite some time ago. On the one hand the original
conditions of adiabatic cyclic unitary evolution were relaxed
quite early \cite{2},\cite{3} and a purely kinematic approach was also 
elaborated \cite{4}. On the other hand the differential geometric framework 
in which the GP is best viewed has been fully delineated \cite{3}-\cite{5} 
- this will be recalled in a specific format below.

As against this situation, the generalisation of the GP concept
from pure states to generic mixed states of quantum systems has
turned out to be non unique, and several different approaches have
been suggested. This is only to be expected as one is 
making a transition from the particular to the general. The approaches
include exploiting the process of purification of a mixed state of a given 
quantum system by tensoring it with another suitably chosen quantum system
and so attaining a pure state \cite{6}; setting up interferometric schemes
in which phase shifts experienced by a system in a mixed state can
be experimentally isolated \cite{7}; using a real metric on the space of
Hilbert-Schmidt operators leading to a natural connection via the
Kaluza-Klein mechanism \cite{8}; and so on.

The purpose of the present work is to approach this problem from a
differential geometric and, in a sense, a minimalist point of
view, including also an essentially unique interpretation based on
the general principles of quantum measurement theory. The main
ingredients are the unitary matrix groups $U(n)$ for general
(unspecified) $n$, some of their coset spaces, and associated
structures. We will first show that the pure state GP problem can
be treated in a systematic way using a set-up involving three
principal fibre bundles (PFB): the first two are specific $U(n)$
coset spaces, the third is an associated bundle (AB) based on the
second. In the second and the third PFB's, the base space
consisting of unitarily related pure state quantum density
matrices is a (co) adjoint orbit in (the dual to) the Lie algebra
${\underline U(n)}$ of $U(n)$. As is well known, such orbits carry
a unique symplectic structure- the Kostant-Kirillov-Souriau (KKS)  symplectic
structure \cite{9}-
\cite{12}-and this is directly related to GP's for
cyclic evolutions. Each of the three PFB's plays a specific role in
the overall picture, with GP's being realised only in the third
one as elements of the $U(1)$ holonomy group. Certain
connections arising naturally in these PFB's will be made use
of, and we will find that the familiar results are immediately
obtained.

The advantage of this set up, which may appear somewhat elaborate
for the pure state case, is that it immediately, easily and
unambiguously generalises to the mixed state situation depending
only upon general quantum principles. One of the important points
we will emphasize is that for cyclic unitary evolutions of such
states, there is no such thing as {\it the} associated GP, but rather
there is a collection of several such phases. However the natural
KKS symplectic structure singles out a specific combination of
them as having a preferred significance, and it is this that can be
directly interpreted along the lines of quantum measurement
theory.

The `minimalist' aspect of the treatment to be given here consists
in the fact that we use only the structures that are already
present in the quantum mechanical description of mixed states.
We merely display them in a particular manner, and then exploit
them to the fullest possible extent. Any other approach, it would
thus appear, must involve ideas and elements in addition to what is
presented here; but in a sense these additions are not really
necessary.

For the pure state GP problem, the case of noncyclic evolutions \cite{3},
the relation to the Bargmann invariants (BI) \cite{4}, uses of geodesics 
\cite{4}, and the more recently discovered Null Phase Curves (NPC) 
\cite{13},\cite{14}  have all been intensively studied. In the present work, 
as we wish to bring out as sharply as possible the most important features of 
mixed state GP's in exclusion to everything else, we shall limit 
ourselves to cyclic evolutions alone. While we will freely use 
geometric and group theoretic ideas intrinsic to the problem, we
will also introduce local coordinate calculations so as to be able
to carry out explicit calculations and make the entire treatment
very tangible.

 The contents of this paper are organised as follows. In Section 2 we
 reformulate the GP associated with pure state unitary cyclic evolution in the
 framework of three PFB's, pointing out the role played by each PFB in the 
overall argument. Section 3 then shows how this framework can be generalised 
in a natural way to evolution of mixed states in the rank two case, leading to 
a physically well defined meaning of the GP to be associated with such cyclic 
evolution. The important role of the KKS symplectic structure in helping us 
identify the mixed state GP is clearly brought out. Section 4 provides the 
physical interpretation of the results of Section 3, bringing in the familiar 
meaning of mixed state density matrices in the context of quantum measurement 
theory. In Section 5 we discuss the role the recently introduced NPC's 
\cite{13},\cite{14} play in the mixed state situation.; this involves
 generalising them  and the associated ideas of Pancharatnam lifts and Null
 Phase Manifolds from pure states to mixed states. The Concluding Section 6 
outlines some general features of the extension of our approach from rank two 
mixed states to higher rank mixed states; contrasts our approach and 
interpretation  with some other treatments; and mentions some open problems.   

\section{Reformulation of pure state GP}
In this section we reformulate the pure state GP using the
framework of coset space PFB's and AB. As explained in the
Introduction we consider only the case of cyclic evolution, as our
main purpose is to extend the treatment to mixed states in later
Sections.

We denote by $\cH$ the Hilbert space of pure states of some
quantum system. We will suppose that $\cH$ is of (complex)
dimension $n$, however in the final GP formulae the parameter $n$
will in fact drop out. The group $U(n)$ of unitary transformations
on $\cH$ will hereafter be denoted by $G$; for the most part we
deal with the defining representation of this group. Its Lie
algebra is described in Appendix A.

The unit sphere in $\cH$ is denoted by $\cB$: 
\be \cB
=\{\psi\in\cH|\;\;||\psi||=1\}\subset \cH , \label{21}
\ee
 and the space of unit rays by $\cR$: 
\be \cR=\cB/U(1)=
\{\rho_\psi=\psi\psi^\dagger |\; \psi\in \cB\} .\label{22}
 \ee
The projection $\pi$ maps $\cB$ onto $\cR$. The preferred or
natural connection one form on $\cB$, whose importance for pure
state GP theory  is well known, is 
\be A=-i\psi^\dagger
d\psi.\label{23} 
\ee 
The two form $dA$ on $\cB$, \be
dA=-id\psi^\dagger\wedge d\psi, \label{24}
\ee 
is the pull back of a symplectic two-form $\Om$ on $\cR$: 
\be dA=\pi^*\Om .\label{25}
\ee 
The intrinsic definition of $\Om$ is as follows \cite{15}.
At each point $\rho\in \cR$, vectors in the tangent space $T_{\rho}\cR$
arise by evaluating the commutators of hermitian operators $K$ on
$\cH$ (generators of $G$) with $\rho$: 
\be \rho\in\cR,\;\; X\in
\;\;T_{\rho}\cR \;\;:\;\; X=-i[K,\rho],\;\;K^\dagger=K.\label{26}
\ee 
Here $K$ is determined by $X$ upto an operator commuting with
$\rho$, but this ambiguity does not matter in the definition of
$\Om$ below. If $\rho=\psi\psi^\dagger$, then a general $X$ and a
$K$ producing it can be expressed in terms of a vector $\chi$
orthogonal to $\psi$ \cite{4}: 
\be
K=i(\chi\psi^\dagger-\psi\chi^\dagger),\;\;
X=\chi\psi^\dagger+\psi\chi^\dagger\;\;, \; \;  (\psi,\chi)=0.
\label{27} 
\ee 
Now $\Om$ is defined at each $\rho$ by giving its
evaluation on two tangent vectors there: 
\bea 
X,\;X^\prime \in T_{\rho}
\cR\;\;: \Om_{\rho}(X,X^\prime) &=&-i{\rm
Tr}(\rho[K,K^\prime])\nonumber\\
&=& 2\;{\rm Im}(\chi,\chi^\prime).\label{28}
\eea
 This $\Om$ is
in fact the Kostant-Kirillov-Souriau (KKS) symplectic two-form on
$\cR$ viewed as a non-generic (co) adjoint orbit in the Lie
algebra ${\underline G}$ of $G$.

The connection $A$ is now used to define horizontal lifts of
smooth curves in $\cR$. If 
\be 
C=\{\rho(s)\in \cR|s_1\leq s\leq
s_2, \; \rho(s_1)=\rho(s_2)\}\subset\cR \label{29}
\ee 
is a
parametrised closed curve in $\cR$, and 
\be 
\cC_h=\{\psi(s)\in
\cB|s_1\leq s\leq s_2 \}\subset\cB \label{210}
\ee 
is a horizontal
lift of $C$ to $\cB$, then at each point of $\cC_h$ we have 
\bea
A_{\psi(s)}(\dot{\psi}(s)) &=& -i(\psi(s),\dot{\psi}(s))\nonumber\\
&=&{\rm Im}(\psi(s),\dot{\psi}(s))=0.\label{211}
\eea 
This lift
$\cC_h$ of $C$ is in general not closed, as $\psi(s_1)$ and
$\psi(s_2)$ may differ by a phase. This is the GP associated with
$C$, and is the $U(1)$ holonomy group element in the sense of
$(\ref{B20})$ in this case: 
\bea
 \label{212} \varphi_{{\rm
geom}}[C]&=& {\rm arg}(\psi(s_1), \psi(s_2))\nonumber\\
&=& -{\int\int}_S\;\Om \;\;\;, \partial S= C, 
\eea where 
$S\in \cR$ is any smooth two-dimensional surface with boundary $C$.

Now we explain the way in which this pure state GP emerges in a
systematic and generalisable manner from a set-up involving three
PFB's, each being used for a particular purpose.

The group $G$ acts transitively on $\cB$. Choose as a 'reference
point' or 'origin' in $\cB$ the first canonical basis vector in
$\cH$, 
\be 
\psi_1^{(0)}=\pmatrix{ 1\cr 0\cr\cdot \cr\cdot\cr
0 } . \label{213}
\ee 
The stability group of $\psi_1^{(0)}$,
namely the subgroup of $G$ leaving $\psi_1^{(0)}$ invariant, is
$H_0=U(n-1)$ acting on dimensions $2,3,\cdots,n$ in $\cH$.
Therefore $\cB$ is the coset space $G/H_0=U(n)/U(n-1)$. The first
coset space PFB we introduce is $(G,\cB,\cdot \cdot,H_0)$, where
for simplicity here and later we omit the symbol for the relevant
projection map. The purpose of this PFB is to help us compute the
Maurer-Cartan one-forms on $G$ in a practically useful form. For a
general $\psi \in \cB$, let $\ell(\psi)$ be some (local) choice of
coset representative, namely an element of $G$ carrying
$\psi_1^{(0)}$ to $\psi$. Therefore $\ell(\psi)$ has the form 
\be
\ell(\psi)=\pmatrix{ ~ & \cdot &\cdots \cdot \cr
                    ~ &\cdot &  \cr
                \psi &\cdot&\cdots \cdot\cr
                    ~ &\cdot &  ~\cr
                    ~ &\cdot &\cdots \cdot } ,
\label{214}
                     \ee
with the first column being $\psi$ and the rest determined upto an
element of $H_0$ on the right. A general matrix $U\in G$ is then
parametrised in the following way: 
\be
U=U(\psi,h_0)=\ell(\psi)h_0,\;\; h_0\in H_0, \label{215}
\ee 
with
$\psi\in \cB$ and $h_0\in H_0$ being (local) coordinates on $G$.
The full set of Maurer-Cartan one-forms on $G$ can now be computed
using eq. $(\ref{B10})$. In the general notation of Appendix B, if
we write the generators of $H_0$ as $J_a$ and the remaining
generators of $G$ as $J_\mu$, eq.$(\ref{B10})$ gives \cite{11},\cite{12}: 
\bea
\label{216} U(\psi,h_0)^{-1}dU(\psi,h_0)&=& -i{\hat
\theta}^{(0)a}J_a - i {\hat \theta}^{(0)\mu}J_\mu\nonumber\\
&=& \psi^\dagger d\psi\;Q_1\;+ {{\underline H}_0}\;{\rm
terms}\;+{\rm cross\;terms},\nonumber\\
Q_1 &=& \pmatrix{ 
1& 0&\cdot &\cdot&0 \cr
                 0&  ~     & ~  & ~ &~\cr
                 \cdot&  ~    &    ~& ~ &~\cr
                 \cdot&   ~   & ~&\mbox{{\LARGE 0}} ~&~\cr
                 0&~ &~ &~ & ~} . 
\eea 
To make contact with the notation of appendix A, the
generators $J_a$ of $H_0$ are $ Q_j,\;J_{jk},\;Q_{jk}$ for
$j,k=2,3,\cdots,n$; the $\psi$-dependent term is the unambiguous
contribution involving the first diagonal generator $Q_1$; and the
cross terms involve $J_{1k},\;Q_{1k}$ for $k=2,3,\cdots,n$, all
outside ${\underline H}_0$. The coefficient of $Q_1$ is independent of the
freedom in the choice of $\ell(\psi)$, and is essentially the one-form $A$ in 
eq.$(\ref{23})$.

Next we turn to the second coset space PFB. The origin
$\psi_1^{(0)}\in \cB$ determines a corresponding point
$\rho^{(0)}= \psi_1^{(0)}\psi_1^{(0)\dagger}\in \cR$. The
stability group of $\rho^{(0)}$ is the subgroup $H=U(1)\times
H_0=U(1)\times U(n-1)\subset G$, $U(1)$ being generated by $Q_1$;
and $\cR$ is the coset space $G/H$. The second coset space PFB is
taken to be $(G,\cR,\cdot \cdot,H)$. Here the base is a particular
(co)adjoint orbit in the Lie algebra ${\underline G}$. On this
PFB, by the definition $(\ref{B33})$, we have a preferred
connection by retaining the terms in eq.$(\ref{216})$ involving
generators of $H$ alone, and dropping the cross terms: 
\bea
\omega^{(2)}&=&-i\left(U(\psi,h_0)^{-1}dU(\psi,h_0)\right)_H\nonumber\\
&=& -i \psi^\dagger d\psi\;Q_1 + {\underline H}_0-{\rm terms}.
\label{217}
\eea

Lastly we bring in a PFB associated to $(G,\cR,\cdot \cdot,H)$:
the base remains the same, while $G$ and $H$ are replaced by
suitably chosen $E$ and $F$. These are: $F=U(1)$ subgroup of
$H=U(1)$ subgroup of $G$ generated by $Q_1$; and $E=\cB$. So this
AB is $(\cB, \cR,\cdot\cdot, U(1))$. The action of $H$ on $F$
which is needed is defined by making $H_0$ in $H$ act trivially, while
$U(1)$ in $H$ acts on $F=U(1)$ by the ( Abelian) $U(1)$ group
composition law. Thus the connection $\omega^{(2)}$ of eq.
$(\ref{217})$ goes over in this third PFB to the connection
\be
\omega^{(3)}= -i \psi^\dagger d\psi = A . \label{218}
\ee

Thus we have arrived at eq. $(\ref{23})$. The ${\underline
H}_0$-terms in $ \omega^{(2)}$ have been dropped since $H_0$ is defined to 
act trivially
on $F=U(1)$, and we have also set $Q_1=1$. In this final result,
the dependence on $n$ and the freedom in the choice of
$\ell(\psi)$ have both disappeared. What we have seen already is
the connection $(\ref{25})$ between $dA$ on $\cB$ and the KKS
symplectic two-form $\Omega$ on $\cR$.

To recapitulate, the first coset space PFB
$(G,\cB,\cdot\cdot,H_0)$ is used along with a choice of coset
representative $\ell(\psi)$ to calculate the Maurer-Cartan
one-forms on $G$ ( at least the terms of interest to us ) in a
convenient manner. This result is used to define a preferred
connection $\omega^{(2)}$ in the second coset space PFB
$(G,\cR,\cdot\cdot,H)$, at the same time bringing in $\cR$ as the
base space. This connection is then 'transferred' to the AB
$(\cB,\cR,\cdot\cdot,U(1))$ and gives back the connection $A$
needed for pure state GP's. In both the second and third PFB's the
base $\cR$ is a (co)adjoint orbit in ${\underline G}$, carrying
the KKS symplectic two-form $\Omega$. In the third PFB, we
recognise that $dA$ on $\cB$ is related to $\Omega$ by pull-back,
and the sequence of operations is complete.

\section{Mixed state GP's}
A pure state density matrix is a rank one operator, with one non
zero eigenvalue unity and the remaining eigenvalues equal to zero.
A mixed state density matrix has in general a spectrum of non zero
eigenvalues each with some multiplicity, followed by a remainder
(in general) of zero eigenvalues. Whereas pure state density matrices are
acted upon transitively by $G$, this is not true for the mixed state case
since both the rank of the density matrix and its spectrum of eigenvalues 
are preserved under unitary transformations. For each rank $k$ the generic
case is when the spectrum of nonzero eigenvalues $\kappa_a$ is non degenerate 
i.e., they obey
\bea
0~<~\kappa_k~<~\kappa_{k-1}<\cdots&&~<\kappa_2~<\kappa_1~<~1,\nonumber\\ 
&& \sum_{a=1}^k~\kappa_a =1.
\label{30}
\eea
The corresponding set of density matrices may be denoted by $
{\cal R}_{\underline \kappa}$. Keeping $k$ and ${\underline \kappa}$ fixed, 
each of these sets is acted 
upon transitively by $G$, and is homeomorphic in a ${\underline \kappa}$- 
dependent manner to the coset space $G/(U(1)^k \times U(n-k))$. Cases of 
degeneracy among the $\kappa_a$ correspond to non generic lower dimensional 
situations described by other coset spaces. 

As the simplest case of a mixed
state we consider rank two density matrices $\rho$ for which the
non zero eigenvalues are non degenerate. Let us write $\kappa_a,
a=1,2$, for these eigenvalues and agree that 
\be
0<\kappa_2<\kappa_1<1 \; ,\;\;\kappa_1+\kappa_2=1. \label{31}\ee 
Then
$\rho$ has the form 
\be \rho=\kappa_1\psi_{1}
\psi_{1}^\dagger+\kappa_2\psi_{2}\psi_{2}^\dagger, \label{32}
\ee
 where the
vectors $\psi_a, a=1,2$, each determined upto a phase factor, form
an ordered orthonormal pair: \be
(\psi_a,\psi_b)=\psi_{a}^\dagger \psi_b=\delta_{ab}. \label{33}
\ee
Hereafter we keep $\kappa_a$ fixed. So each such $\rho$ is in unique
one to one correspondence with an ordered pair of pure state
density matrices defined as and obeying 
\bea
\rho_a&=&\psi_{a}\psi_{a}^\dagger\; , \; \rho_a\rho_b= \delta_{ab}\rho_a \;\;
({\rm no\;sums!}),\nonumber\\
\rho&=& \kappa_1\rho_1+\kappa_2\rho_2. \label{34}
\eea 
This set of
$\rho$'s forms a (co)adjoint orbit under $G$. At the vector space
level we have to deal with ordered pairs $\psi_a, a=1,2$, as in
eqs.$(\ref{32},\ref{33})$. We recognise here the generalisations of
$\cB$ and $\cR$ of the pure state situation to mixed states of the
form $(\ref{32})$, in which for any ${\underline \kappa}=(\kappa_1, 
\kappa_2)$ with $\kappa_2~<~\kappa_1$ and $\kappa_1+\kappa_2=1$, we have an
orbit ${\cal R}_{{\underline \kappa}}^{(2)}$ replacing ${\cal R}$. 
 We now define and describe these spaces in detail, stressing that we need 
something at the vector space level 'on top of' density matrices. 
\vskip0.3cm 
\noindent
{\underline{The space ${\cB}^{(2)}$}}\\

We define this space to consist of ordered pairs of orthonormal
vectors in $\cH$, with no explicit mention of $\kappa_a$. For
later convenience we write the pair of vectors in a particular
notation: 
\be {\cB}^{(2)}=\{\Psi=(\psi_1\;\;\psi_2)|\psi_a \in
\cB, \psi_a^\dagger\psi_b=\delta_{ab}\}. \label{35}
\ee 
In an
obvious manner, the group $G$ acts transitively on ${\cB}^{(2)}$.
A convenient 'origin' consists of the first two canonical basis
vectors in $\cH$: 
\bea
\Psi^{(0)}&=&(\psi_{1}^{(0)}\;\;\psi_{2}^{(0)})\nonumber\\
\psi_1^{(0)}=\pmatrix{ 1\cr 0\cr\cdot \cr \cdot\cr 0 }&,&
\psi_2^{(0)}=\pmatrix{ 0\cr 1\cr\cdot \cr\cdot\cr 0 }.
\label{36} 
\eea 
The stability group of $\Psi^{(0)}$ is the
subgroup $H_0=U(n-2)\subset G$ acting on the dimensions
$3,4,\cdots,n$ in $\cH$. (The use of the same letter $H_0$, and
later $H$, as in the previous section should cause no confusion.)
Thus we recognise ${\cB}^{(2)}$, the orbit of $\Psi^{(0)}$ under $G$
action, as a coset space: 
\bea
{\cB}^{(2)}&=&G/H_0=U(n)/U(n-2),\nonumber\\
{\rm dim}{\cB}^{(2)}&=&4(n-1).\label{37}
\eea 
Elements of the
tangent space to ${\cB}^{(2)}$ at $\Psi$ can be described as
follows. Each $\Phi \in T_{\Psi}{\cB}^{(2)}$ is a pair
$\Phi=(\phi_{1}\;\;\phi_{2})$, $\phi_a\in\cH$, obeying
restrictions which follow from eq.$(\ref{33})$: 
\bea
(\psi_1,\phi_1), (\psi_2,\phi_2)&=& {\rm pure
~imaginary},\nonumber\\
(\psi_1,\phi_2)+(\phi_1,\psi_2)&=&0. \label{38}
\eea 
Taking out a
factor of $i$ we can write each such $\Phi$ uniquely as 
\bea
\Phi&=& i\Psi h + {\overline {\underline \chi}},\nonumber\\
h^\dagger&=&h={\rm 2\times 2~~matrix},\nonumber\\
 {\overline {\underline \chi}}&=&(\chi_1\;\;\chi_2),\nonumber\\
 \chi_a&\in& {\cH}_\perp (\Psi)={\rm subspace~ of~ \cH~ orthogonal~ to
 ~\psi_1 ~and ~\psi_2}. \label{39}
\eea
Thus we have a one to one correspondence 
\be 
\Phi \in T_\Psi
{\cB}^{(2)} \leftrightarrow h,\;{\overline {\underline
\chi}}.\label{310}
\ee 
This is a generalisation of the pure state
case where any $\phi\in T_{\psi}\cB$ has the unique form \cite{4}
 \bea
\phi&=&ia\psi+\chi,\nonumber\\a^*&=&a,\nonumber\\\chi &\in&
\cH_{\perp}(\psi) . \label{311}
\eea 
The real number $a$ gets generalised
to a $2\times 2$ hermitian matrix $h$, while $\chi\in
\cH_{\perp}(\psi)$ has been replaced by an ordered pair $
{\overline {\underline \chi}}=(\chi_1\;\;\chi_2)$ with each
$\chi_a\in \cH_{\perp}(\Psi)$.

\vskip0.3cm \noindent 
{\underline{The space ${\cR}^{(2)}$}}\\

 This is the space of mixed state density matrices we are interested in,
and it can be described in several useful ways: 
\bea 
{\cR}^{(2)}
&=&\{\rho^\dagger=\rho \geq 0, {\rm Tr}\rho =1|{\rm Spectrum\; of\;
}\rho = (\kappa_1,\kappa_2,0,\cdots,0)\}\nonumber\\
&=&\{U\rho^{(0)}U^{-1}|U\in G, \rho^{(0)}=\kappa_1\psi_{1}^{(0)}
\psi_{1}^{(0)\dagger}+\kappa_2\psi_{2}^{(0)}
\psi_{2}^{(0)\dagger}\}\nonumber\\
&=& \{(\rho_1 \; \;  \rho_2)|\rho_a\in \cR, \rho_1\rho_2=0\}.
\label{312}
\eea
The last description of ${\cR}^{(2)}$ ( we omit ${\underline \kappa}$ in 
 ${\cR}_{{\underline \kappa}}^{(2)}$ since ${\underline \kappa}$ is kept 
fixed in the discussion), in which $\kappa_a$ do not appear explicitly, 
is actually equivalent to the earlier description, via a 
${\underline \kappa}$- dependent diffeomorphism. However we do not mention
this repeatedly. 

Under $G$ action, the stability group of $\rho^{(0)}$
is $H=U(1)\times U(1) \times H_0$, the $U(1)$ factors acting on
the first and the second directions in $\cH$. Thus we exhibit
$\cR^{(2)}$ as a coset space which is in fact a (co)adjoint orbit
in ${\underline G}$, as well as a quotient space starting from
$\cB^{(2)}$: 
\bea 
\cR^{(2)}&=& {\rm (co)\; adjoint\; orbit\; of}
\;\rho^{(0)}\nonumber\\
&=& G/H \nonumber\\
&=&\cB^{(2)}/U(1)\times U(1), \nonumber\\
{\rm dim}\cR^{(2)} &=&{\rm dim}\cB^{(2)}-2=2(2n-3). \label{313}
\eea 
The (${\underline \kappa}$ dependent ) projection $\pi:\cB^{(2)}\rightarrow 
\cR^{(2)}$ takes
$\Psi\in \cB^{(2)}$ to $\rho_{\Psi}\in \cR^{(2)}$ according to 
\bea
\rho_\Psi &=& \pi(\Psi)=\Psi \kappa \Psi^\dagger,\nonumber\\
\kappa &=&\pmatrix{ \kappa_1&0\cr 0&\kappa_2 } .\label{314} 
\eea 

The description of the tangent spaces $T_\rho\cR^{(2)}$ involves a
little effort. If we use the representation $(\ref{314})$ for
$\rho_\Psi$, and take some $\Phi \in T_\Psi \cB^{(2)}$, a general
$X\in T_{\rho_\Psi}\cR^{(2)}$ is certainly expressible as 
\be
X=\Psi \kappa \Phi^\dagger+\Phi \kappa \Psi^\dagger\in {\underline G}.
\label{315}
\ee 
Using eq.$(\ref{39})$ for $\Phi$ and writing out
and grouping terms, we see that 
\be 
X=i\Psi[h, \kappa]\Psi^\dagger+\Psi
\kappa {\underline {\overline \chi}}^\dagger + {\underline {\overline
\chi}}\kappa\Psi^\dagger . \label{316} 
\ee 
This is certainly determined
by $h$ and ${\underline {\overline \chi}}$, but $h_{11}$ and
$h_{22}$ are not needed since 
\bea
[h,\kappa ]&=&(\kappa_1-\kappa_2)\pmatrix{0&-h_{12}\cr
h_{21}&0 },\nonumber\\h_{21}&=&h_{12}^* . \label{317}
\eea
Therefore, as is easily confirmed, each $X\in T_{\rho_{\Psi}}
\cR^{(2)}$ is determined by, and corresponds in a one-to-one
fashion to, a complex number $h_{12}$ and a pair $\chii$: 
\be
X(h_{12},\chii)=-i(\kappa_1-\kappa_2)(h_{12}\psi_1\psi_{2}^\dagger
-h_{12}^*\psi_2\psi_{1}^\dagger)+
\kappa_1(\psi_1\chi_{1}^\dagger+\chi_{1}\psi_{1}^\dagger)
+\kappa_2(\psi_{2}\chi_{2}^\dagger+\chi_{2}\psi_{2}^\dagger) .
\label{318}
\ee 
If we alter $\psi_a$ by independent phases
$e^{i\alpha_a}$ which leave $\rho_{\Psi}$ invariant, to keep $X$
unchanged we must replace $h_{12}\rightarrow
e^{i(\alpha_2-\alpha_1)}h_{12}, \chi_a \rightarrow e^{i\alpha_a}
\chi_a$. Returning to $\Phi \in T_{\Psi} \cB^{(2)}$ in
$(\ref{39})$, we can tentatively separate it into vertical and
horizontal parts, the former being the $h_{11}, h_{22}$ terms and
the latter the rest: 
\bea 
\label{319} \Phi &=&i\Psi h + \chii \\
&=& i \Psi\pmatrix { h_{11}&0\cr 0&h_{22} } 
+i\Psi \pmatrix {0&h_{12}\cr h_{12}^*&0 } + \chii . \nonumber
\eea 
The horizontal part is in unambiguous correspondence with $X$ in eq.
$(\ref{318})$.

For the later determination of the KKS two-form on $\cR^{(2)}$, we
need to express each $X\in T_{\rho_\Psi}\cR^{(2)}$ as the
commutator of some hermitian operator $K\in {\underline G}$ with
$\rho_\Psi$. This is easily done: 
\bea
X(h_{12},\chii)&=&-i[K(h_{12},\chii), \rho_\Psi]\nonumber , \\
K(h_{12},\chii)&=&
i(\chii\Psi^\dagger-\Psi\chii^\dagger)-\Psi \pmatrix{0&h_{12}\cr
h_{12}^*&0  }  \Psi^\dagger.\label{320}
\eea 
The presence of new terms compared to eq.$(\ref{27})$ in the pure state case 
should be noted.

We may add the following remark. Each (co)adjoint orbit (fixed by 
${\underline \kappa}$ as explained above ) meets the subalgebra of 
diagonal matrices in as many points as the number of diagonal matrices 
we get by applying the permutation group ( Weyl group ) to the starting 
diagonal matrix $ \rho^{(0)} = \kappa_1 \psi_{1}^{(0)}\psi_{1}^{(0)\dagger}
+ \kappa_2 \psi_{2}^{(0)}\psi_{2}^{(0)\dagger} + \cdots 
\kappa_k \psi_{k}^{(0)}\psi_{k}^{(0)\dagger}$, just intersecting each 
Weyl chamber exactly once. Fixing ${\underline \kappa}$ in such a way that 
$0~<~\kappa_k~<~\kappa_{k-1}<\cdots~<\kappa_2~<\kappa_1~<~1$ is then 
equivalent to choosing a particular Weyl chamber. Therefore, we have as many
orbits as the points in the interior of a Weyl chamber, the boundary points 
corresponding to the case where the mixed state density matrix has degenerate 
eigenvalues. For example, in the rank two case, analysed explicitly in this
Section, the Weyl chamber is a one-dimensional segment that we have chosen to 
parametrize by $\kappa_1\in [1/2,1)$. 
\vskip0.3cm \noindent 
{\underline{Local coordinates on
${\cB}^{(2)}$and ${\cR}^{(2)}$}}\\

To later connect Hilbert space notations with differential
geometric ones, we now describe correlated local coordinate
choices around general points in ${\cB}^{(2)}$ and in
${\cR}^{(2)}$. Take a point $\Psi_0=(\psi_{01}\;\;\psi_{02}) \in
{\cB}^{(2)}$, not necessarily the 'origin' $\Psi^{(0)}$ of eq
$(\ref{36})$. Its image in ${\cR}^{(2)}$ is 
\be 
\Psi_0\in
{\cB}^{(2)}\rightarrow \rho_0=\pi(\Psi_0)=\Psi_0 \kappa \Psi_{0}^\dagger
\in {\cR}^{(2)} .\label{321}
\ee 
Convenient neighbourhoods of
$\Psi_0, \rho_0$ will get determined as we describe them. The
orthogonal complement to $\Psi_0$, a subspace of $\cH$ of complex
dimension $n-2$, is defined as 
\be 
\cH_\perp(\Psi_0)=\{\psi\in
\cH|(\psi_{0a},\psi)=0,\;a=1,2\}\subset \cH . \label{322}
\ee
 Let
$\Psi = (\psi_1\;\;\psi_2) \in {\cB}^{(2)}$ be 'near' $\Psi_0$.
Then each of $\psi_1$ and $\psi_2$ is expressible as a unique linear 
combination of
$\psi_{01},\psi_{02}$ plus some vector in $\cH_\perp(\Psi_0)$. Let
us write 
\bea 
\psi_a &=&S_{ba}\psi_{0b}+\chi_{0a},
a=1,2,\nonumber\\
\chi_{0a}&&\in \cH_\perp(\Psi_0),\nonumber\\
{\rm i.e.} \Psi&=&\Psi_0\;S +\chii_0, \label{323}
\eea 
with $S$ a complex $2\times 2$ matrix. The condition $(\ref{33})$ becomes:
\be 
S^\dagger S= 1_{2\times 2} - \chii_{0}^\dagger \chii_0 .
\label{324}
\ee 
Let us then limit $\chii_0$ so that the two eigenvalues
of $\chii_{0}^\dagger \chii_0 =(\chi_{0a}^\dagger \chi_{0b})$ both
lie in $[0,1)$. (This means $\chii_0$ involves $4(n-2)$ real
independent variables.) This makes $S$ non singular, the general
solution being 
\be S=\cU(1-\chii_{0}^\dagger \chii_0)^{1/2},\;
\cU\in U(2).\label{325}
\ee 
Here the square root is the unique
hermitian positive definite one, so this is the polar
decomposition of $S$.

If we allow $\cU$ to be a general $U(2)$ element, that brings in
four new independent variables, so $\cU$ and $\chii_0$ together
account for 4(n-1) real independent variables which would be right
for $\cB^{(2)}$. However the action of $U(1)\times U(1)$ on $\Psi$
amounting to a motion along fibres is 
\be
\Psi \rightarrow \Psi \pmatrix{
e^{i\alpha_1}&0\cr 0&e^{i\alpha_2}}, \label{326}
\ee 
and it
is convenient to have the charts on $\cB^{(2)}$ and $\cR^{(2)}$
related in this way. We therefore limit $\cU$  in eq.$(\ref{325})$
to a two parameter family. We see easily that if $\cU\in U(2)$ has
real positive diagonal elements, then it is actually an element of
$SU(2)$ and takes the form 
\be 
\cU(z)=\pmatrix{\sqrt{1-|z|^2}&
z\cr -z^*&\sqrt{1-|z|^2}}  \;\;,|z|<1. \label{327}
\ee 
We thus have a
local coordinate description of a neighbourhood of $\rho_0$ in
$\cR^{(2)}$ as follows: a point $\rho \in \cR^{(2)}$ near $\rho_0$
is 
\bea \rho&=&\Psi \kappa \Psi^\dagger,\nonumber\\
\Psi&=& \Psi_0\cU(z)(1-\chii_{0}^\dagger\chii_0)^{1/2} + \chii_0.
\label{328}
\eea 
In all, $z$ and $\chii_0$ amount to $2(2n-3)$ real
independent parameters, the dimension of $\cR^{(2)}$. The
neighbourhood of $\rho_0$ is defined by the conditions on
$\chii_0$ and $z$ in eqs. $(\ref{325},\ref{327})$. For each $\rho$
in this neighbourhood, we have a unique lift $\Psi\in {\cB}^{(2)}$ given in
$(\ref{328})$. A general $\Psi^\prime \in \pi^{-1}(\rho)$ differs
from $\Psi$ by a diagonal phase matrix : 
\be 
\Psi^\prime =\Psi \pmatrix{
e^{i\alpha_1}&0\cr 0 & e^{i\alpha_2} } , \;\; 0\leq
\alpha_1, \alpha_2 < 2\pi. \label{329}
\ee 
Both $\rho$ and $\Psi$
in eq $(\ref{328})$ are functions of $z$ and $\chii_0$. In
addition $\Psi^\prime$ involves $\alpha_1$ and $\alpha_2$. At
$\rho_0$ and $\Psi_0$ both $z$ and $\chii_0$  vanish. At
$\Psi_0$, $\alpha_1=\alpha_2=0$ as well. To compare eqs.$(\ref{328})$, 
$(\ref{329})$ with the pure state case, see \cite{14}.

\vskip0.3cm \noindent 
{\underline{Vectors and forms at $\Psi_0$}}\\

Since the matrix $(1-\chii_{0}^\dagger \chii_{0})^{1/2}$ is not
easy to differentiate, we limit ourselves to small regions in $
\cR^{(2)}$ and $\cB^{(2)}$ around $\rho_0$ and $\Psi_0$
respectively. By $\Phi^\prime$ we denote a general tangent
vector in $T_{\Psi^\prime}\cB^{(2)}$ in the manner of
eq.$(\ref{39})$. We will actually need expressions for
$X_{\Phi^\prime}$, $A^{(a)}$ and $dA^{(a)}$ (defined later) at
$\Psi_0$, which means we ultimately take $\Phi^\prime \in
T_{\Psi_0}\cB^{(2)}$. For these purposes we find that it is
adequate to retain only terms linear in $\alpha_a,z$ and
$\chii_0$. From eqs. $(\ref{328},\ref{329})$ we have: 
\bea
\psi_{1}^\prime
&=&\psi_{01}(1+i\alpha_1)-z^*\psi_{02}+\chi_{01},\nonumber\\
\psi_{2}^\prime
&=&\psi_{02}(1+i\alpha_2)+z\psi_{01}+\chi_{02}.\label{330} 
\eea
Next we let $\Phi^\prime \in T_{\Psi_0}\cB^{(2)}$ correspond to
the pair $h^\prime, \chii^\prime$ in the sense of $(\ref{39})$.
Then the nearby point $\Psi^\prime=\Psi_0+\epsilon\Phi^\prime $,
for small $\epsilon$, involves small changes
$\delta\alpha_1,\delta\alpha_2,\delta z, \delta\chi_{0a}$ around
zero, obtained by comparison with eq. $(\ref{330})$: 
\bea
\Psi^\prime &=& \Psi_0+\epsilon\Phi^\prime:\nonumber\\
&&\delta\alpha_1=\epsilon
h_{11}^\prime,\;\;\delta\alpha_2=\epsilon
h_{22}^\prime,\nonumber\\
&&\delta z=i\epsilon h_{12}^\prime,\;\; \delta z^*=-i\epsilon
h_{21}^\prime,\nonumber\\
&& \delta\chi_{01}=\epsilon \chi_{1}^\prime,\;\;
\delta\chi_{02}=\epsilon \chi_{2}^\prime . \label{331} 
\eea 
Dropping
$\epsilon$, the standard differential geometric way of
representing $\Phi^\prime$ at $\Psi_0$ is as  
\bea 
X_{\Phi^\prime}
&=& h_{11}^\prime \frac{\partial}{\partial \alpha_1}+
h_{22}^\prime \frac{\partial}{\partial \alpha_2}+i h_{12}^\prime
\frac{\partial}{\partial z}- i h_{21}^\prime
\frac{\partial}{\partial
 z^*} \nonumber\\&+& \frac{\partial}{\partial \chi_{01}} \chi_{1}^\prime
+\frac{\partial}{\partial \chi_{02}} \chi_{2}^\prime + \chi_{1}^
 {\prime\dagger}\frac{\partial}{\partial \chi_{01}^\dagger}+
\chi_{2}^{\prime\dagger}\frac{\partial}{\partial
\chi_{02}^\dagger} .\label{332}
\eea
In a similar spirit we compute $A^{(a)}$ and $dA^{(a)}$ at
$\Psi_0$. For the former we find 
\be A^{(a)} =-i\psi_{a}^{\prime
\dagger}d\psi_{a}^{\prime}=d\alpha_a ,\label{333}
\ee 
which implies
\be 
i_{X_{\Phi^\prime}} A^{(a)} =h_{aa}^\prime, \;\;
a=1,2. \label{334}
\ee 
Thus as anticipated in eq.$(\ref{319})$ 
\be
\Phi^\prime\;{\rm horizontal} \Leftrightarrow i_{X_{\Phi^\prime}}
A^{(a)}=0 \Leftrightarrow h_{11}^\prime=h_{22}^\prime =0.
\label{335}
\ee 
Now we look at the two-forms $dA^{(a)}$ again at
$\Psi_0$. Simple calculations give the results 
\bea 
dA^{(1)}&=&
-idz\wedge dz^* -i d\chi_{01}^\dagger \wedge d\chi_{01},\nonumber\\
dA^{(2)}&=& +idz\wedge dz^*- i d\chi_{02}^\dagger \wedge
d\chi_{02} .\label{336}
\eea 
We can contract these with tangent
vectors $\Phi^\prime, \Phi^{\prime\prime}$ using eq.$(\ref{332})$
and we then get 
\bea
dA^{(1)}(X_{\Phi^\prime},X_{\Phi^{\prime\prime}}) &=&i_{X_{\Phi\prime\prime}}
i_{X_{\Phi^\prime}}\;dA^{(1)}\nonumber \\
&=&-i(h_{12}^\prime h_{21}^{\prime\prime}-h_{21}^\prime
h_{12}^{\prime\prime})+i(\chi_{1}^{\prime\prime\dagger}
\chi_{1}^{\prime}-
\chi_{1}^{\prime\dagger}\chi_{1}^{\prime\prime}),\nonumber\\
dA^{(2)}(X_{\Phi^\prime},X_{\Phi^{\prime\prime}}) &=&i_{X_{\Phi\prime\prime}}
i_{X_{\Phi^\prime}}\;dA^{(2)}\nonumber \\
&=&+i(h_{12}^\prime h_{21}^{\prime\prime}-h_{21}^\prime
h_{12}^{\prime\prime})+i(\chi_{2}^{\prime\prime\dagger}
\chi_{2}^{\prime}-
\chi_{2}^{\prime\dagger}\chi_{2}^{\prime\prime}). \label{337}
\eea
With these preparations we can go on to GP considerations.

\vskip0.3cm \noindent {\underline{The PFB framework and GP's}}\\

We now follow the same pattern of arguments as in the previous
Section for pure states. The first coset space PFB is now
$(G,\cB^{(2)},\cdot\cdot,H_0)$ with $H_0=U(n-2)$. A choice of
coset representative at $\Psi \in \cB^{(2)}$ is of the form 
\bea
\ell(\Psi)&=&\pmatrix{ ~ & \cdot &\cdots \cdot \cr
                    ~ &\cdot &  \cr
                \Psi &\cdot&\cdots \cdot\cr
                    ~ &\cdot &  ~\cr
                    ~ &\cdot &\cdots \cdot}
                    \in G,\nonumber\\
                     \ell(\Psi)\Psi^{(0)}&=& \Psi. \label{338}
\eea
This replaces eq.$(\ref{214})$, and $\ell(\Psi)$ is arbitrary upto
an element of $H_0$ on the right. A general matrix $U\in G$ is
parametrised as 
\be 
U(\Psi,h_0)=\ell(\Psi)h_0, \:\:h_0\in H_0
\label{339}
\ee 
in place of $(\ref{215})$. The replacement for eq.
$(\ref{216})$ involving all the Maurer-Cartan forms on $G$ is 
\be
U(\Psi,h_0)^{-1}dU(\Psi,h_0)= \psi_{1}^\dagger d\psi_{1}\;Q_1\;+
\psi_{2}^\dagger d\psi_{2}\;Q_2\;+ {{\underline H}_0}\;{\rm
terms}\;+{\rm cross\;terms}.\label{340}
\ee

The second coset space PFB is $(G,\cR^{(2)},\cdot\cdot,H)$ with
$H=U(1)\times U(1)\times H_0$. The preferred connection on this PFB
is obtained from eq.$(\ref{340})$ by dropping the cross terms and
retaining only the $H$-terms: 
\bea
\omega^{(2)}&=&-i\left(U(\Psi,h_0)^{-1}dU(\Psi,h_0)\right)_H\nonumber\\
&=& -i \psi_{1}^\dagger d\psi_{1}\;Q_1 -i \psi_{2}^\dagger
d\psi_{2}\;Q_2  + {\underline H}_0-{\rm terms}. \label{341}
\eea
which replaces eq.$(\ref{217})$.

The third PFB is an AB to the previous one in which we replace $G$
and $H$ by suitable $E$ and $F$: $E=\cB^{(2)}$, $F=U(1)\times
U(1)$ part of $H$. The action of $H$ on $F$ is defined again by
making $H_0$ act trivially, while $U(1)\times U(1)$ acts on $F$
following the abelian composition law. Thus from $\omega^{(2)}$ we
arrive at the connection 
\be \omega^{(3)}= -i \psi_{1}^\dagger
d\psi_{1}\;Q_1 -i \psi_{2}^\dagger d\psi_{2}\;Q_2 \label{342}  ,
\ee
on this third PFB. Now we cannot delete $Q_1$ and $Q_2$ here as
they are the two independent generators of the two $U(1)$ factors
in $U(1)\times U(1)$. Alternatively we can say we have two
independent one-forms $A^{(a)}$ on $\cB^{(2)}$: 
\be
A^{(a)}=-i\psi_{a}^\dagger d\psi_{a}\;\; {\rm (no\;sum)}.
\label{343}
\ee 
while the $\underline{U(1)}\times\underline{ U(1)}$ valued connection
$\omega^{(3)}$ is 
\be 
\omega^{(3)}= A^{(1)}\;Q_1 +A^{(2)}\;Q_2  . 
\label{344} 
\ee 
The evaluations of $A^{(a)}$ and $dA^{(a)}$ on
tangent vectors at general points on $\cB^{(2)}$ are contained in
eq. $(\ref{334},\ref{337})$.

If we consider a closed curve $C\subset \cR^{(2)}$ (cyclic mixed
state evolution), a horizontal lift $\cC_h\subset \cB^{(2)}$ must
obey two conditions at each point: 
\bea
A_{\Psi(s)}^{(a)}(\dot{\Psi}(s)) &=&0, \nonumber\\
{\rm i.e.,~ } (\psi_a(s), \dot{\psi}_a(s))&=&0,\;\;a=1,2. \label{345}
\eea 
In
general now the end points of $\cC_h$ differ by a pair of phases,
an element of $U(1)\times U(1)$, not just by a single phase. Each
of them is a GP and should be counted independently. This leads us
to consider the two independent two-forms $dA^{(a)}$ on
$\cB^{(2)}$. On the other hand, the KKS construction leads to a
single symplectic two-form $\Omega$ on $\cR^{(2)}$, so the
question is to find out which linear combination of $dA^{(a)}$ is
related to $\Omega$ via pullback. We now find this combination.

\vskip0.3cm \noindent {\underline{The KKS two-form on
$\cR^{(2)}$}}\\

In eq.$(\ref{320})$ we have an expression for a general tangent
vector $X \in T_\rho \cR^{(2)}$, as well as a hermitian generator
$K$ leading to it upon commutation with $\rho$. The KKS symplectic
two-form $\Omega$ on $\cR^{(2)}$ is defined at each point by its
evaluation on two tangent vectors \cite{15}: 
\be 
\Omega_\rho(X^\prime,
X^{\prime\prime}) =-i {\rm Tr}_\cH (\rho\;[K^\prime,
K^{\prime\prime}])  . \label{346}
\ee 
For clarity we have indicated that 
the trace has to be computed on the Hilbert space $\cH$. Using
eq.$(\ref{320})$ we find after some algebra: 
\bea
\Omega_\rho(X^\prime,
X^{\prime\prime})&=&-i(\kappa_1-\kappa_2)(h_{12}^\prime
h_{21}^{\prime\prime}-h_{21}^\prime
h_{12}^{\prime\prime})\nonumber\\
&-&i\kappa_1(\chi_{1}^{\prime\dagger} \chi_{1}^{\prime\prime}-
\chi_{1}^{\prime\prime\dagger}\chi_{1}^{\prime})
-i\kappa_2(\chi_{2}^{\prime\dagger} \chi_{2}^{\prime\prime}-
\chi_{2}^{\prime\prime\dagger}\chi_{2}^{\prime}) .\label{347}
\eea
Comparing this with the expressions for
$dA^{(a)}(X_{\Phi^\prime},X_{\Phi^{\prime\prime}})$ in
eq.$(\ref{337})$  we see that we have the relation 
\be
\sum_{a}\kappa_a dA^{(a)} = \pi^* \Omega  . \label{348}
\ee 
Here
finally the non zero eigenvalues $\kappa_a$ of $\rho \in
\cR^{(2)}$ have reappeared, and at the same time dependences on
$n$ have disappeared.

 This approach indicates that the unique GP we can associate with
 a cyclic evolution in the coadjoint orbit of a given rank $2$ mixed state 
density operator is a linear combination of the two phases provided by the
 $U(1)\times U(1)$ holonomy group element , and this combination
 is expressible as the symplectic area of a surface in
 $\cR^{(2)}$:
\bea  
\varphi_{{\rm
geom}}^{(a)}[C]&=& {\rm arg}(\psi_a(s_1), \psi_a(s_2)) \; , \; a=1,2 \; ; 
\nonumber\\
\sum_{a} \kappa_a\varphi_{{\rm geom}}^{(a)}[C] &=&
-{\int\int}_S\;\Om \;\;\;, \partial S= C .\label{349} 
\eea Here
$C=\{\rho(s)\}$ is a closed loop on $\cR^{(2)}$ and
$\cC_h=\{\Psi(s)\}$ is a horizontal lift of it in $\cB^{(2)}$.

We explore the physical interpretation of these results in the
next Section.

\section{Physical interpretation of mixed state GP's}

The present approach to mixed state unitary evolution based on the
PFB framework has naturally emphasized the fact that ( in the rank
two case) the holonomy group is $U(1)\times U(1)$ . So at the end
of a cyclic evolution we have a pair of geometric phases
$\varphi_{{\rm geom}}^{(a)}[C]$, not simply one. On the other hand
the KKS definition of a canonical symplectic structure on the
space of these density matrices, which form a (co)adjoint orbit in
${\underline G}$, leads to a unique two-form $\Omega$ given in eqs.
$(\ref{346},\ref{347})$. The symplectic area integral of $\Omega$
is a weighted average of the two GP's, as in eq. $(\ref{349})$. We
now construct an interpretation of this result, based on general
quantum mechanical principles.

A mixed state density matrix $\rho$ for a quantum system is a
convex combination of any number of pure state density matrices \cite{16}:
\be 
\rho= \sum_{r} p_r \rho_r,\;\;\rho_r\in \cR,\;\; p_r>0,\;
\sum_{r}p_r = 1. \label{41}
\ee 
(Of course there must be at least
two terms present). Here the $p_r$ are any set of classical
probabilities. The $\rho_r$ do not have to be pairwise orthogonal.
A mixed $\rho$ can be expanded in this form in infinitely many
ways, and each expansion represents a distinct physical way in
which an ensemble of kinematically identical systems,
characterised as a whole by $\rho$, can be synthesised. Given the
particular expansion $(\ref{41})$, we can imagine an ensemble of a
very large number of systems, a fraction $p_r$ of which form a sub
ensemble in the pure state $\rho_r$. The average of the results of
measurements of any hermitian observable $\theta$ over the entire
ensemble is given by 
\be 
<\theta>=\sum_{r} p_r {\rm
Tr}(\rho_r\theta)= {\rm Tr}(\rho\theta). \label{42}
\ee 
In the
final result only $\rho$ appears, not the particular way in which
the ensemble was physically prepared. This expresses the physical
fact that the average of measurements over any one of these
ensemble realisations of $\rho$ is always the same. Of course,
${\rm Tr}(\rho\theta)$ need not be any one of the eigenvalues of
$\theta$; even each individual ${\rm Tr}(\rho_r\theta)$ need not
be an eigenvalue of $\theta$.

Among the infinitely many realisations $(\ref{41})$ of $\rho$ is
of course a special or canonical one. This corresponds to the
spectral resolution of $\rho$ when the $p_r$ are the non zero
eigenvalues $\kappa_a$ of $\rho$ (assumed non degenerate for
simplicity), and the $\rho_a$ are the corresponding mutually
orthogonal pure state projections. ( In this case, the number of
terms in eq.$(\ref{41})$ cannot exceed ${\rm dim}\; \cH =n$). Our
result $(\ref{349})$ for mixed state GP's suggests that we use
this canonical ensemble realisation of $\rho$.

We now go back to the rank two case and use the canonical
decomposition $(\ref{32})$. The measurement of GP's is not like the
measurement of some hermitian operator observable belonging to the
system under consideration. Let us nevertheless imagine that we
have an ensemble of systems, a fraction $\kappa_1$ of which are in
the pure state $\rho_1=\psi_{1}^\dagger \psi_1$, and the remaining
fraction $\kappa_2$ are in the orthogonal pure  state $\rho_2=\psi_{2}^\dagger
\psi_2$. As $\rho$ undergoes unitary cyclic evolution, so do each
of $\rho_1$ and $\rho_2$, but these latter are pure state
evolutions. We assume an experimental arrangement has been set up
which is capable of measuring these two pure state GP's. Then the
ensemble average of the results of these measurements is exactly
what appears in eq.$(\ref{349})$ on the left hand side, which need
not be the same as either of the two individual $GP's$ ( or indeed
any GP). However this ensemble averaged GP is what is reproduced
by the symplectic area calculation on $\cR^{(2)}$, using the
canonical KKS two-form $\Omega$.

This 'minimalist' interpretation works only with the canonical
ensemble realisation of $\rho$, and involves an average of phases,
not of unimodular phase factors $\exp(i\varphi_{\rm
geom}^{(a)}[C])$. This implies that the experimental measurements
of the $\varphi_{\rm geom}^{(a)}[C]$ must not be just modulo
$2\pi$, but must keep careful track of the gradually accumulating value
of each $\varphi_{\rm geom}^{(a)}[C]$ as the cyclic evolution is
experienced.

\section{ The relation of geometric phase to null phase curves for mixed 
states}

In this Section we would like to generalize some earlier results on Berry's 
phase for pure states \cite{14}. In particular, we would like to show 
how geometric phase(s), for both cyclic and noncyclic evolutions,  can be 
directly obtained as a surface integral of the KKS symplectic two-form once a 
suitable class of curves, the null phase curves, has been defined. For 
definiteness, we will consider again the case of rank two density matrices, 
but the results can be easily generalized to the higher rank situation which 
will be briefly described in the last Section. 

Let us start by recalling some of the geometrical structures we have studied 
in the preceding sections. We have seen that, for each $\underline{\kappa}= 
(\kappa_1,\kappa_2), \; \kappa_1+\kappa_2=1$, the space  
$\cR^{(2)}_{\underline{\kappa}}$ can be identified with the adjoint orbit 
under the $U(n)$ action of a given rank-two density matrix  $\rho^{(0)} = 
\kappa_1 \rho_1^{(0)}  + \kappa_2 \rho_2^{(0)} $. This orbit is, in turn, 
isomorphic  to the coset space $U(n)/U(1)\times U(1)\times U(n-2)$, via  a 
$\underline{\kappa}$-dependent map which fixes  the two $U(1)$'s  as being 
generated by $\rho_j ^{(0)} $ , $j=1,2$. On each orbit $\{ \rho= U \rho^{(0)} 
U^{-1} = \kappa_1 U\rho_1^{(0)}U^{-1}  + \kappa_2 U\rho_2^{(0)}U^{-1} \, | \, 
U\in U(n)\}$, the KKS symplectic form is given by $\Omega_{\rho}  = \mbox{Tr} 
(\rho\, dU^\dagger \wedge dU )$ and its pull-back to the full $U(n)$ as well 
as to the bundle space $\cB^{(2)}$ is exact, with 
$\pi_{\underline{\kappa}}^\star \Omega_{\rho} = dA_{\rho}$, $A_{\rho} = 
\kappa_1 A_1 + \kappa_2 A_2= \mbox{Tr}( \rho\, U^\dagger dU)$. We notice that 
both $\Omega_{\rho}$ and $A_{\rho}$ depend explicitly on  
$\underline{\kappa}$ i.e. they are specific to the chosen orbit, on which 
we confine the evolution to define geometric phases. For this reason, in the 
following we drop the subscript $\underline{\kappa}$.

We consider continuous parametrized curves ${\cal C} \in \cB^{(2)}$ and their 
projections to $C= \pi ({\cal C}) \subset \cR^{(2)}$:
\bea
&&{\cal C} = \{ ( \psi_1(s) ~ \psi_2(s) ) \in \cB^{(2)} \, | \, s\in [s_1,s_2] 
\} \\
&& C = \{ \rho(s) = \kappa_1 \rho_1(s) + \kappa_2 \rho_2(s) \in \cR^{(2)} \, | 
\,  s\in [s_1,s_2] \} 
\eea
with the following smoothness conditions:\\
the curves ${\cal C}$, $C$,  are said to be ${\textit class~ I~ curves}$  iff 
$\psi_j(s), \rho_j(s)$ are continuous, piecewise differentiable and
\be
(\psi_j(s_1),\psi_j(s_2)) \neq 0,\;  j=1,2;
\ee
the curves ${\cal C}$, $C$,  are said to be ${\textit class~ II~ curves}$ iff 
$\psi_j(s), \rho_j(s)$ are continuous, once differentiable and
\be
(\psi_j(s),\psi_j(s')) \neq 0, \; j=1,2, \mbox{ for any } s,s'\in[s_1,s_2].
\ee
In addition, a curve ${\cal C}$, $C$ of class II is said to be a 
{\it null phase curve} (NPC) iff 
\be
\mbox{Tr}(\rho_j(s) \rho_j(s^{\prime}) \rho_j(s^{\prime\prime})) =\mbox{real 
positive} \, \Leftrightarrow \,  
\mbox{Tr}(\rho_j(s) [\rho_j(s^{\prime}), \rho_j(s^{\prime\prime})]) = 0 , \; 
j=1,2, \; \mbox{for any } s,s^{\prime},s^{\prime\prime}\in [s_1,s_2]. 
\label{npc} 
\ee 
We can understand this definition also from a more geometrical point of view. 
Let us consider the subset of couples of vectors $( \psi_1 ~ \psi_2 ) 
\in \cB^{(2)}$ such that $\psi_j$ ($j=1,2$) belongs to the real linear hull 
obtained by forming all real linear combinations of any number of vectors 
$\psi_j(s^\prime)$ (renormalized if necessary). This collection of couples is 
associated to a real subspace of ${\cal H}^{(2)} = 
\{ ( \psi_1 ~ \psi_2 ) \, | \, \psi_j \in {\cal H}\}$ which, because of 
(\ref{npc}), is $\pi^\star \Omega$ isotropic.  We are thus led to characterize 
a NPC via such associated subspaces. 

Given a class II curve $C = \{ \rho(s) = \kappa_1 \rho_1(s) + \kappa_2 
\rho_2(s)\}  \in \cR^{(2)}$, we can define its {\it Pancharatnam lift} to a 
curve ${\cal C}_0 = \{ ( \psi_1^0(s) ~ \psi_2^0(s) ) \} \in \cB^{(2)}$ such 
that, for each component:
\be 
(\psi_j^0(s), \psi_j^0(s^{\prime}))= \mbox{real positive for any } s,s' \in 
[s_1,s_2], \label{pl}
\ee
in a way similar to  the construction obtained in \cite{14} for the pure 
state case. Choosing any reference point $( \psi_1^0(s_0) ~ \psi_2^0(s_0) ) 
\in \cB^{(2)}$, this lift is explicitly determined by setting, for $j=1,2$:
\bea
\psi_j^0(s) &=& N_j(s) \rho_j(s) \psi_j^0(s_0) ,\\
N_j(s) &=& |(\psi_j^0(s_0), \psi_j(s) )|^{-1} = 
[ \mbox{Tr}(\rho_j^0 \rho_j)]^{-1/2}.
\eea

As a consequence of (\ref{pl}), any two points of ${\cal C}_0$ are in phase 
in the Pancharatnam sense and the curve ${\cal C}_0$ is horizontal:
\bea
\arg(\psi^0_j(s_{1}), \psi_j^0(s_{2}))  &  = &0,\nonumber\\
\int_{\mathcal{C}_{0}} \;A_j  &  = & 0 ,
\eea
where $A_j = -i \psi_j^\dagger  d\psi_j$. It is then not difficult to check 
that, for a general lift $\cC=\{ (e^{i\alpha_1(s)} \psi^0_1(s) ~ 
e^{i\alpha_2(s)} \psi^0_2(s))\}$ of $C$ obtained from $\cC_0$ by a smooth 
local $U(1)\times U(1)$ phase transformation, one has:
\be 
\int_{\mathcal{C}}\;A_j  = \int_{s_1}^{s_2} ds \frac{d\alpha_j(s)}{ds} = 
\alpha_j(s_2)-\alpha_j(s_1) = \arg(\psi_j(s_{1}),\psi_j(s_{2})) . \label{prop}
\ee

We are now ready to define the geometric phase (GP) associated to any class I
curve $C = \{ \rho(s)\}$ from $\rho(s_1)$ to $\rho(s_2)$. Let  $C'$ be any NPC 
from $\rho(s_2)$ to $\rho(s_1)$ so that $C\cup C^{\prime}$ is a class I closed 
loop. Then, if $S$ is a two-dimensional surface  such that $\partial 
S=C\cup C^{\prime}$, the GP associated to $C$ is defined to be given by:
\be
\varphi_g[C]    = - \;\int_{S}  \Omega \label{gp}.
\ee
With some algebra, one can easily show that the integral (\ref{gp}) is indeed 
independent of the choice of the NPC $C'$ and that the geometric phase 
associated to any NPC vanishes. Also the  kinematic definition of the GP is 
recovered: if ${\cal C}$ is any lift of $C$, from $(\psi_1(s_1)~\psi_2(s_1)) 
$ to $ (\psi_1(s_2)~\psi_2(s_2))$, one has  
\bea
\varphi_g[C] &\equiv & -\int_S \Omega =  -\oint_{\cC \cup \cC'} A = -
\int_{\cC} A - \int_{\cC'} A =  \nonumber \\
&=&  \arg(\psi_1(s_1),\psi_1(s_2)) +  \arg(\psi_2(s_1),\psi_2(s_2))  - 
\int_{\cC} A.
\eea
 
There are additional properties of GP's that are worth mentioning and that 
can be recovered from definition (\ref{gp}) and from the property (\ref{prop}) 
of  NPC's. Suppose first that $C_{12}, C_{23}, C_{31}$ are projections of the 
NPC's
$\cC_{12}, \cC_{23}, \cC_{31}$ from $(\psi_1(s_1)~\psi_2(s_1)) $ to $ 
(\psi_1(s_2)~\psi_2(s_2))$, from $(\psi_1(s_2)~\psi_2(s_2)) $ to $ 
(\psi_1(s_3)~\psi_2(s_3 ))$ and from $(\psi_1(s_3)~\psi_2(s_3)) $ to $ 
(\psi_1(s_1)~\psi_2(s_1))$ respectively. Since both $C_{12}\cup C_{23}\cup  
C_{31}$ and $\cC_{12}\cup  \cC_{23} \cup \cC_{31}$ are closed loops, we have
\bea
\varphi_g[C_{12}\cup C_{23}\cup  C_{31}] &=& - \oint_{\cC_{12}\cup 
\cC_{23}\cup  \cC_{31}}  A
= -\oint_{\cC_{12}} A -\oint_{\cC_{23}} A -\oint_{\cC_{31}} A \nonumber \\
&=& - \kappa_1 \, \mbox{argTr} (\rho_1(s_1) \rho_1(s_2) \rho_1(s_3)) - 
\kappa_2 \, \mbox{argTr} (\rho_2(s_1) \rho_2(s_2) \rho_2(s_3)) . 
\eea
More generally, for any class I curves $C_{12}, C_{23}, C_{31}$ which are 
projections of $\cC_{12}, \cC_{23}, \cC_{31}$ we can prove the relation:
\be
\varphi_g[C_{12}\cup C_{23}\cup  C_{31}]  = \varphi_g[C_{12}] + 
\varphi_g[C_{23}] + \varphi_g[C_{31}] - \kappa_1 \mbox{arg Tr} (\rho_1(s_1) 
\rho_1(s_2) \rho_1(s_3)) - \kappa_2 \mbox{arg Tr} (\rho_2(s_1) \rho_2(s_2) 
\rho_2(s_3)) ,
\label{noadd}
\ee
showing the lack of additivity of the GP. 

Let us now consider a connected, simply connected smooth submanifold $M \in 
\cR^{(2)}$with dimension $m\geq2$ in the real sense and let us denote with 
$\iota_M : M \hookrightarrow  \cR^{(2)}$ the corresponding inclusion map. By 
using eq. (\ref{noadd}) above, one can show that if $M$ is a Null Phase 
Manifold (NPM), i.e a submanifold  such that every once-differentiable curve 
$C \subset M$ is a NPC, then:
\bea
&& \mbox{ M is isotropic: }  \Omega_M\equiv \iota_M^\star\Omega =0  ; 
\label{iso}\\
&& \mbox{ for any } \rho= \pi( (\psi_1~\psi_2)) , \rho^\prime= 
\pi( (\psi_1^\prime~\psi_2^\prime)) , \rho^{\prime \prime} 
= \pi( (\psi_1^{\prime \prime}~\psi_2^{\prime \prime}))\in M , \nonumber \\
&& ~~~~~~~~~~~~ \mbox{Tr} (\rho_1 \rho_1^\prime\rho_1^{\prime \prime}), 
 \mbox{Tr} (\rho_2 \rho_2^\prime\rho_2^{\prime \prime}) \mbox{ are real 
positive} .\label{bar}
 \eea
Let us first concentrate on (\ref{iso}), which shows that isotropy is  a 
necessary condition for $M$ to be a NPM. We will see now that it is not a 
sufficient one. To examine this point, let us suppose that $M$ is such that 
$\mbox{Tr}(\rho_1 \rho_1^\prime) > 0$, $\mbox{Tr}(\rho_2 \rho_2^\prime) > 0$ 
for any $\rho=\kappa_1 \rho_1 + \kappa_2 \rho_2, \; \rho^\prime = \kappa_1 
\rho_1^\prime + \kappa_2 \rho_2^\prime$. In the spirit of the Pancharatnam 
lift defined in eq. (\ref{pl}), we can construct a lift of $M$ to a 
submanifold $M_0 \in \cB^{(2)}$ as follows. Given a point $\rho \in M$, its 
lifted point  $(\psi_1~\psi_2) \in \cB^{(2)}$ is given by the choice:
\be
\psi_1= \frac{\rho_1 \psi_1^0}{\sqrt{\mbox{Tr}(\rho_1^0 \rho^1)}} \; , \; 
 \psi_2= \frac{\rho_2 \psi_2^0}{\sqrt{\mbox{Tr}(\rho_2^0 \rho^2)}}.
\ee
where $\rho^0$, $(\psi_1^0~ \psi_2^0)$ are fiducial points in $M$, $M_0$ 
respectively, and  $\pi( (\psi_1^0~ \psi_2^0)) = \rho^0 $. This lift is 
characterized by the fact that any point $(\psi_1~\psi_2) \in M_0$ is in 
phase with $(\psi_1^0~ \psi_2^0)$ in the Pancharatnam sense:
\be
(\psi_1^0, \psi_1), (\psi_2^0, \psi_2) > 0.
\ee
In general, however, two generic points $(\psi_1~ \psi_2), (\psi_1\prime~ 
\psi_2^\prime)\in M_0$ are not in phase since:
\be
(\psi_j^\prime,\psi_j) = \mbox{Tr}(\rho_j^0 \rho_j^\prime \rho_j), \; j=1,2 .
\ee
If now we suppose $M$ to be isotropic, one can easily prove that, for any two 
class I curves in $M$  from $\rho(s_1)$ to $\rho(s_2)$, say $C_{12}$ and 
$ C_{12}^\prime$, one has:
\be
\varphi_g[ C_{12}] = \varphi_g[ C_{12}^\prime] ,
\ee
i.e., denoting with $\cC_{12},  \cC_{12}^\prime$ the corresponding lifts in 
$M_0$:
\be
\int_{\cC_{12}} A = \int_{\cC_{12}^\prime } A.
\ee
This means that the pull-back of $A$ from $\cB^{(2)}$ to $M_0$ is exact. 
Thus, setting  $\iota_{M_0} : M_0 \hookrightarrow \cB^{(2)}$, we have the 
result:
\be
\Omega_M = 0 \; \Leftrightarrow \; \iota_{M_0}^\star A = df .
\ee
If in addition $M$ is a NPM we have the stronger result:
\be
 \iota_{M_0}^\star A = 0 ,
\ee
which follows from the fact that now $ (\psi_j^\prime,\psi_j)>0$, $j=1,2$, for 
any two points in $M_0$. This result gives the extent to which the 
NPM property goes beyond isotropy.

To find a sufficient condition for $M$ to be a NPM one has to consider 
(\ref{bar}). One can finally assert the following inverse result 
\cite{14}: if $M$ is such that for any three points 
$\rho= \pi( (\psi_1~\psi_2)) , \rho^\prime= 
\pi( (\psi_1^\prime~\psi_2^\prime)) , \rho^{\prime \prime} = 
\pi( (\psi_1^{\prime \prime}~\psi_2^{\prime \prime}))$, the quantities 
$\mbox{Tr} (\rho_1 \rho_1^\prime\rho_1^{\prime \prime}), 
 \mbox{Tr} (\rho_2 \rho_2^\prime\rho_2^{\prime \prime})$ are real positive, 
then:
\bea
&& \mbox{Tr}(\rho_1\rho_1^\prime),\; \mbox{Tr}(\rho_2\rho_2^\prime) >0 ;\\ 
&&  M \mbox{ is an NPM};\\
&&  M \mbox{ is isotropic}.
 \eea
Notice that these three statements are not independent, since the third is 
implied by the second. 
 
\section{Concluding Remarks}

We have set up what may be called a 'minimalist' interpretation
for the meaning to be given to the phrase 'mixed state GP',
limiting ourselves for clarity to the case of unitary cyclic
evolutions. We have been guided by the structures of, and
relationships among, certain PFB's which arise naturally in this
context. They all flow out of the unitary group $G=U(n)$ acting on
the $n$-dimensional Hilbert space of a quantum system. Our aim has
been to bring into focus the role of the KKS symplectic structure
existing on each (co)adjoint orbit in ${\underline G}$. In the
final results, as often stated, explicit dependences on $n$
actually drop out. This is because in these results only the codimensions 
are relevant.

We considered the case of rank two density matrices $\rho$, with
the two non zero eigenvalues obeying $0<\kappa_2< \kappa_1 <1$. It
can be seen fairly easily that the framework set up in this paper,
involving three PFB's in sequence and the use to which each is
put, can be faithfully repeated for higher rank ( but still non
degenerate for non zero eigenvalues ) density matrices. The main
features for rank $k,\; 0<k<n$, would be that the non zero
eigenvalues of $\rho$ would obey 
\be
0<\kappa_k<\kappa_{k-1}<\cdots <\kappa_2<\kappa_1<1,\;\;
\sum_{a=1}^{k}\kappa_a=1. \label{51}
\ee 
Then $\rho$ has the decomposition 
\be \rho = \sum_{a=1}^{k} \kappa_a \psi_a
\psi_a^\dagger, \;\; (\psi_a, \psi_b)=\delta_{ab}. \label{52}
\ee
The stability groups $H_0$ and $H$ in this situation would be
$H_0=U(n-k)$ acting on dimensions $(k+1), (k+2),\cdots , n$ of
$\cH$; $H=U(1)\times U(1)\times \cdots\times U(1)\times H_0$, with
$k$ $U(1)$ factors. Correspondingly at the vector and operator
levels we have to deal with the spaces 
\bea \cB^{(k)} &=&
\{\Psi=(\psi_1 \;\;\psi_2\;\;\cdots \psi_k)|\psi_a\in \cB,
(\psi_a,\psi_b)=\delta_{ab}\}\nonumber\\
&=&G/H_0,\nonumber\\
\cR^{(k)}&=& \{\rho=\sum_{a=1}^{k} \kappa_a\rho_a| \rho_a=
\psi_a \psi_{a}^{\dagger}\in \cR\}\nonumber\\
&=&G/H\nonumber\\
&=&\cB^{(k)}/U(1)\times U(1)\times \cdots\times U(1). \label{53}
\eea 
These spaces are of real dimensions $k(2n-k)$  and
$k(2n-k-1)$ respectively, and the latter is always even, with
$\cR^{(k)}$ being a (co)adjoint orbit in ${\underline G}$.

The sequence of three PFB's is now
$(G,\cB^{(k)},\cdot\cdot,H_0),\;(G,\cR^{(k)},\cdot\cdot,H)$ and
$(\cB^{(k)},\cR^{(k)},\cdot\cdot,U(1)\times U(1)\times
\cdots\times U(1))$. On the last we obtain, following the set up
given earlier, the connection one-form 
\bea
\omega^{(3)}&=&\sum_{a=1}^{k} A^{(a)} Q_a,\nonumber\\
A^{(a)}&=&-i \psi_a^\dagger d\psi_{a}.\label{54}
\eea 
This serves to
define the concept of horizontal lifts of a curve
$C\subset\cR^{(k)}$ to $\cC \subset\cB^{(k)}$. The KKS symplectic
two-form $\Omega$ on $\cR^{(k)}$ is however unique, and its
relation to the above $A^{(a)}$ is 
\be 
\sum_{a=1}^{k}\kappa_a
dA^{(a)}=\pi^*\Omega. \label{55}
\ee

The general interpretation follows lines similar to what is
described in Sections 4 and 5 . As the holonomy group is $U(1)\times
U(1)\times \cdots\times U(1)$ ($k$ factors), a cyclic evolution of
such mixed states naturally involves $k$ separate $U(1)$ phases or
$k$ separate pure state GP's $\varphi_{\rm geom}^{(a)}[C]$. What
the KKS structure does is to relate a particular linear
combination of these to a two dimensional symplectic area integral
in $\cR^{(k)}$.

For emphasis, we may restate our results in the following intuitive 
manner. Consider the case of rank $n$ (maximal rank) non degenerate density
matrices $\rho$, belonging to $\cR^{(n)}$  and with eigenvalues arranged in
decreasing order $\kappa_1,~\kappa_2,\cdots,~\kappa_n$. Such a $\rho$
determines an orthonormal basis or frame in Hilbert space upto $n$ phases, 
namely upto an element of $ U(1)\times\cdots\times U(1)$ ($n$ factors). Given 
a closed trajectory ( cyclic unitary evolution) of the density martrix 
in $\cR^{(n)}$, the different possible unitary evolutions which will carry the
density matrix along the given trajectory will differ from one  another at
each point by independent$ U(1)\times\cdots\times U(1)$ phases. The 
$ U(1)\times\cdots\times U(1)$ relative phases at the level of $\cB^{(n)}$ 
between the final and the initial frames have two parts: a dynamical part 
depending on the particular unitary evolution chosen, and one that depends
only on the closed trajectory in $\cR^{(n)}$. Then the available invariant 
or geometric quantities that remain are an $n$-tuple of $U(1)$ abelian
phases. Any function of these is also a geometric invariant. Our analysis of
the canonical KKS symplectic structure on $\cR^{(n)}$ singles out a particular
such function as having a preferred significance. 

The considerations of \cite{7} have certain points of similarity with the
above. The concept of horizontal lift of an evolution in $\cR^{(k)}$
to one in $\cB^{(k)}$ is similar; in our treatment explicit use is made of 
the third PFB $(\cB^{(k)},~\cR^{(k)},\cdots,U(1)\times\cdots\times U(1))$ and
the connection $\omega^{(3)}$ of eq.$(\ref{54})$ thereon. However, while our 
framework of three PFB's seems to play no explicit role in ref \cite{7}, the
use of the KKS symplectic structure on $\cR^{(k)}$ above gives a satisfying 
underpinning to arrive at the weighted sum of geometric phases tied to the
spectral decomposition of $\rho$.     

The concept of off-diagonal GP's for multi ($n$) level quantum
systems has been recently introduced and studied in the
literature \cite{17}, \cite{18}. Here too for such systems we have $n$ 
individual pure
state GP's defined for generic unitary cyclic evolution, and in
addition several algebraically independent Bargmann invariants (of
order four) also enter the picture. The spirit of the present
paper has some points of similarity with off-diagonal GP ideas.

In case we have degenerate mixed states, in the sense that some
non zero eigenvalues of $\rho$ have non trivial multiplicity, we
have to deal with non Abelian holonomy groups \cite{19}, rather than just
products of $U(1)$ factors. This would naturally lead us to non
Abelian GP's, but the basic three-PFB scheme set up here would
again be available.

\vskip1cm

\newpage
\def\theequation{A.\arabic{equation}}
\centerline{\appendix{\bf{APPENDIX  A: THE LIE ALGEBRA OF  $U(n)$}}}
\setcounter{equation}{0}
\vskip1cm

 The family of compact unitary groups
$U(n)$ plays an important role in our analysis. Mainly for
notational convenience we list the generators and commutation
relations in the defining representation.

The definition of $U(n)$ is 
\be U(n)=\{U= n\times n {\rm
complex\;matrix}|U^\dagger U =1_{n\times n}\}.\label{A1}
\ee
Regarding this as a group of complex rotations in an
$n$-dimensional complex space, subgroups $U(n-1), U(n-2), \cdots,
U(1)$ can easily be identified in various ways.

The generators of this defining representation of $U(n)$ consist
of all $n\times n$ hermitian matrices. These may be separated into
pure imaginary antisymmetric matrices $J_{jk}=-J_{kj}$, generating
the $SO(n)$ subgroup of $U(n)$, and real symmetric 'quadrupole'
matrices $Q_{jk}= Q_{kj}$; here the indices $j,k$ go over the
range $1, 2, \cdots,n$. The definitions are 
\bea (J_{jk})_{\ell
m}&=&\frac{i}{\sqrt{2}}(\delta_{j\ell}\delta_{km}-\delta_{jm}\delta_{k\ell})
,\nonumber \\
(Q_{jk})_{\ell
m}&=&\frac{1}{\sqrt{2}}(\delta_{j\ell}\delta_{km}+\delta_{jm}\delta_{k\ell}).
\label{A2}
\eea 
Their commutation relations separate into three
sets: 
\bea -i[J_{jk}, J_{\ell m}]&=&
\frac{1}{\sqrt{2}}(\delta_{k\ell}J_{jm}-\delta_{j\ell}J_{km}+\delta_{km}J_{\ell
j}- \delta_{jm}J_{\ell k}), \nonumber\\
-i[J_{jk}, Q_{\ell m}]&=&
\frac{1}{\sqrt{2}}(\delta_{k\ell}Q_{jm}-\delta_{j\ell}Q_{km}+\delta_{km}Q_{j
\ell}- \delta_{jm}Q_{k\ell }), \nonumber\\
-i[Q_{jk}, Q_{\ell m}]&=&
\frac{1}{\sqrt{2}}(\delta_{k\ell}J_{mj}+\delta_{j\ell}J_{mk}+\delta_{km}J_{\ell
j}+ \delta_{jm}J_{\ell k}) . \label{A3}
\eea

The numerical coefficients in eq.$(\ref{A2})$ have been chosen so
that as far as possible these matrices are trace orthonormal. We
have: 
\bea 
{\rm Tr}(J_{jk} J_{\ell
  m})&=&\delta_{j\ell}\delta_{km}-\delta_{jm}\delta_{kl},
\nonumber\\
{\rm Tr}(J_{jk} Q_{\ell m})&=& 0, \nonumber\\
{\rm Tr}(Q_{jk} Q_{\ell
  m})&=&\delta_{j\ell}\delta_{km}+\delta_{jm}\delta_{kl}.
\label{A4}
\eea
Thus while distinct generators are definitely 'trace orthogonal',
each individual $J_{jk}$ and each individual $Q_{jk}$ for $j\neq
k$ have normalised traces in the above sense. The exceptional
cases are the generators $Q_{11}, Q_{22},\cdots, Q_{nn}$ since for
$j=k=\ell=m$ we have a factor of $2$ on the right hand side in the
last of eqs.$(\ref{A4})$. To have a strictly trace orthonormal
basis for the Lie algebra ${\underline U(n)}$ of $U(n)$ in the
defining representation we therefore may take the basis to be made
up of 
\bea J_{jk}&=&-J_{kj},\nonumber\\
Q_{jk}&=&Q_{kj},\;\;j\neq k, \nonumber\\
Q_{j}&=&\sqrt{2}Q_{jj}\;\;{\rm no\;sum\;on}\;j.\label{A5}
\eea 
Each
of the matrices $ J_{jk}$ and $Q_{jk}$ for $j\neq k$ has no non
vanishing diagonal matrix elements. On the other hand, each
$Q_{j}$ has a matrix element of unity at the $j^{th}$ place in the
diagonal, while all other matrix elements vanish. Using the basis
$(\ref{A5})$ we can write a general element of ${\underline U(n)}$
as a real linear combination of the form: 
\bea 
X&=& x_j Q_j
+\frac{1}{2}x_{jk}J_{jk}+\frac{1}{2}x_{jk}^\prime Q_{jk},
\nonumber\\
&&x_{jk}=-x_{kj}\;\;x_{jk}^\prime= x_{kj}^\prime\;\;{\rm for}\;
j\neq k.\label{A6}
\eea 
Then we have the trace formula 
\be 
{\rm Tr}(XY)=x_{j}y_{j}+\frac{1}{2}x_{jk}y_{jk}+\frac{1}{2}x_{jk}^\prime
y_{jk}^\prime, \label{A7}
\ee 
so each independent term appears with a coefficient of unity.

Regarding ${\underline U(n)}$ as the abstract Lie algebra of
$U(n)$ we may denote its basis elements  corresponding to the
above matrices as 
\be 
Q_j\rightarrow e_j\;\;,J_{jk}\rightarrow
e_{jk}=-e_{kj}\;\;,Q_{jk}\rightarrow e_{jk}^\prime =e_{kj}^\prime
\;\;{\rm for}\;j\neq k . \label{A8}
\ee 
The $Q_j$, or $e_j$ in a
general situation, are the generators of the Abelian torus
subgroup $U(1)\times U(1)\times \cdots \times U(1)$ of $U(n)$,
consisting of all diagonal matrices $U$.

\vskip1cm
\def\theequation{B.\arabic{equation}}
\centerline{\bf{APPENDIX B: PRINCIPAL FIBRE BUNDLES, ASSOCIATED BUNDLES,}} 
\centerline{\bf{COSET SPACES AND CONNECTIONS}}
\setcounter{equation}{0}
\vskip1cm

 For setting notations and as a ready reference for the reader, we here
 collect briefly the basic definitions and properties of the structures named 
above \cite{20}-\cite{23}.

\vskip0.3cm \noindent {\underline{Principal fibre bundles and
connections }\\

A principal fibre bundle (PFB) is a collection of four objects written as 
$(P,M,\pi,H)$: $P$ the total space; $M$ the base space; $\pi$ the projection 
map $P\rightarrow M$; and $H$ a Lie group, the structure group and typical 
fibre. $P$, $M$ and $H$ are all differentiable manifolds, with ${\rm dim}\; 
P={\rm dim}\; M+{\rm dim}\; H $. Points in them will be denoted by
$p,p^\prime,\cdots,\;\;m,m^\prime,\cdots,\;\;h,h^\prime,\cdots,$ with 
$\pi(p)=m,\;\;\pi(p^\prime)=m^\prime,\;\cdots $. For each element $h\in H$ 
there is a globally well defined fibre-preserving diffeomorphisn $\psi_h$ of 
$P$ onto itself, which is free and transitive on each fibre. In a local 
trivialization of the bundle, the portion $\pi^{-1}(M_\alpha) \subset P$ 
lying 'on top of' some open subset $M_\alpha \subset M$ 'looks like' the 
Cartesian product $M_\alpha \times H$. We express this with the compact 
notation 
\be 
\pi(p)=m\in M_\alpha\;\;:\;\;p=(m,h)_\alpha,\;\;h\in
H, \label{B1}
\ee 
$h$ being uniquely determined by $p$. As $h$ varies over $H$ with $m$ kept 
fixed, we obtain all points $p\in \pi^{-1}(m)$. In the overlap of two such 
local trivializations we have a transition rule 
\bea 
\pi(p)=m\in M_\alpha \cap
M_\beta: && p=(m,h)_\alpha=(m,h^\prime)_\beta :\nonumber\\
&& h^\prime=t_{\beta\alpha}(m)h,\;\; t_{\beta\alpha}(m) \in H,\label{B2}
\eea 
with the transition group element appearing by convention on the left hand 
side. In the relevant overlaps these transition functions obey 
\bea
t_{\alpha\beta}(m)^{-1}&=&t_{\beta\alpha}(m),\nonumber\\
t_{\alpha\beta}(m)t_{\beta\gamma}(m)&=&t_{\alpha\gamma}(m).\label{B3}
\eea
The globally well defined map $\psi_{h^\prime}$ representing $h^\prime \in H$ 
appears locally (by convention, so as not to 'interfere' with the transition 
rule $(\ref{B2})$) as a right translation along fibres: 
\be
\psi_{h^\prime}(m,h)_\alpha=(m,hh^{\prime -1})_\alpha.
\label{B4}
\ee 
In this set up, we do not contemplate any action of $H$ on $M$.

A connection on $P$ is a one-form $\omega$ on $P$ taking values in the Lie 
algebra ${\underline H}$ of $H$, and obeying two important conditions spelt 
out below. We denote by $e_a$ the elements of a basis for ${\underline H}$, 
so ${\underline H}={\rm Sp}\{e_a\}$. At each point $p\in P$, the tangent 
space $T_p P$ contains a vertical subspace $V_p$ corresponding to motions 
within the fibre induced by the (right) actions $\psi_h$ of elements $h\in
H$. This leads to a natural isomorphism $\rho_{p}:V_p \rightarrow 
{\underline H}$ with $\rho_{p}^{-1}: H\rightarrow V_p$. The first condition 
on $\omega$ is that at each $p$, the contraction of vertical vectors with 
$\omega_p$ should agree with $\rho_p$:
\be 
v\in V_p:i_v\omega_p=\rho_p(v)\in \underline{H} \; , \label{B5}
\ee 
the second is the 'equivariance' condition which controls the behaviour of 
$\omega_p$ as $p$ runs over a fibre: 
\be 
h\in H: \psi_{h}^*\omega = \cD(h^{-1})\circ \omega,\label{B6}
\ee 
where $\cD(h)$ is the adjoint representation of $H$ on $\underline{H}$, with
matrices (${\cD^a}_b(h)$). One can show that in a local trivialization of $P$
over $M_\alpha \subset M$, $\omega$ necessarily has the form \cite{23}: 
\be 
m\in M_\alpha, p=(m,h)_\alpha :
\omega_p = (\hat{\Theta}^{(0)a}(h)-{\cD^a}_b(h^{-1})A^{b}(m))e_a,
A^a \in {\mathcal X}^*(M_\alpha).\label{B7}
\ee 
Here $\hat{\Theta}^{(0)a}$ are the left-invariant Maurer-Cartan one-forms on 
$H$, adapted to the basis $\{e_a\}$ for ${\underline H}$; and each $A^a$ is 
a one-form defined locally over $M_\alpha$. ( We omit the extra label $\alpha$ 
on these one-forms). If $m$ is in the overlap $M_\alpha \cap M_\beta$ of the 
domains of two local trivializations of $P$, then the two expressions for 
$\omega$ involving $A^a$ over $M_\alpha$ and $A^{\prime a}$ over $M_\beta$ 
are related by the gauge transformation formula 
\be A^{\prime
a}(m) ={\cD^a}_b(t_{\beta \alpha}(m)) (A^
b(m)+\hat{\Theta}^{(0)b}(t_{\beta \alpha}(m))).\label{B8}
\ee 
Any ${\underline H}$-valued one-form $\omega$ on $P$ obeying the two 
conditions $(\ref{B5},\ref{B6})$, described locally as in $(\ref{B7})$ 
subject to the transition rule $(\ref{B8})$, is an acceptable connection on 
$P$, there being no preferred one.

Sometimes for practical calculations it is convenient to work within some 
(unspecified) matrix representation $\cU(h)$ of $H$, with generators $e_a 
\rightarrow -iJ_a$. Then eqs.$(\ref{B7},\ref{B8})$ have the convenient matrix 
forms 
\bea
\omega_p&=&\cU(h)^{-1}(id-J_aA^a(m))\cU(h),\nonumber\\
J_aA^{\prime a}(m)&=&\cU(t_{\beta\alpha}(m))(J_aA^a(m)-id)
\cU(t_{\beta\alpha}(m))^{-1}, \label{B9}
\eea 
while the Maurer-Cartan one-forms appearing in eqs.$(\ref{B7},\ref{B8})$ are 
obtainable from 
\be
\cU(h)^{-1}d\cU(h)=-i\hat{\Theta}^{(0)a}J_a.\label{B10}
\ee

Given a connection $\omega$ on $P$, at any $p\in P$ the horizontal subspace 
$H_p \subset T_p P$ is defined to be the null space of $\omega_p$: 
\be 
X \in T_p P: X\in H_p \Leftrightarrow
i_X\omega_p=0. \label{B11}
\ee 
Then $T_pP$ appears as the direct
sum of vertical and horizontal subspaces:
\be 
T_pP=V_p\oplus H_p,
\label{B12}
\ee 
and the tangent map $(\pi_*)_p$, which annihilates
$V_p$, gives an one-to-one onto map of $H_p$ to $T_mM$ in the base:
\be
\pi(p)=m:(\pi_*)_p: V_p\rightarrow 0, H_p\rightarrow T_mM.
\label{B13}
\ee

The last item in this brief recapitulation of PFB structure is the concept of 
parallel transport, or horizontal lift of a smooth curve in $M$ upto $P$. 
Let $C=\{m(s)\}\subset M$ be a smooth parametrised curve in the base. 
Then a smooth parametrised curve $\cC=\{p(s)\}\subset P$ is a horizontal 
lift of $C$ (with respect to a given connection $\omega$) if $\cC$ projects 
onto $C$ and at each point its tangent vector is horizontal: 
\bea \pi(p(s))&=&
m(s),\nonumber\\
X(s)&=& {\rm tangent\;to} \;\cC \; {\rm at}\; p(s) \in
H_{p(s)}.\label{B14}
\eea

In local coordinates, say $q^\mu$ for $M$ and $\theta^a$ for $H$, along with
an accompanying local trivialization of $P$, we get explicit formulae suitable
for computations. The entire set $(q^\mu,\theta^a)$ gives a local coordinate
system for $P$. The Maurer-Cartan one forms $\hat{\Theta}^{(0)a}$ and the
one-forms $A^a$ determining $\omega$ may be written as 
\bea
\hat{\Theta}^{(0)a}(h)&=& \hat{\xi}_{b}^{a}(\theta)  d\theta^b,\nonumber \\ 
A^a(m)&=& A_{\mu}^{a}(q) dq^\mu. \label{B15}
\eea 
Let us write the coordinates of points on $C$ as $q^\mu (s)$; for a horizontal 
lift $\cC$ we must determine the additional coordinates $\theta^a (s)$ such 
that condition $(\ref{B14})$ is obeyed. The tangents to $C$ and $\cC$ at 
corresponding points are 
\bea 
\frac{dq^\mu(s)}{ds} \frac{\partial}{\partial q^\mu} &&\in T_{m(s)}
M,\nonumber\\
X(s)=\frac{dq^\mu (s)}{ds} \frac{\partial}{\partial q^\mu} &+&
\frac{d\theta^a(s)}{ds} \frac{\partial}{\partial \theta^a} \in T_{p(s)} P. 
\label{B16}
\eea 
Then $(\ref{B14})$ becomes a system of first order ordinary differential 
equations for the coordinates $\theta^a(s)$ of a variable element $h(s)\in H$: 
\be
\hat{\xi}_{b}^{a}(\theta(s)) \frac{d\theta^b
(s)}{ds}={\cD^{a}}_b(h(s)^{-1})A_{\mu}^{b}(q(s))\frac{dq^\mu
(s)}{ds}.\label{B17}
\ee 
In a general matrix representation this
takes the form 
\be 
\frac{d}{ds}\cU(h(s)) =-iJ_a
A_{\mu}^a(q(s))\frac{dq^\mu (s)}{ds}\cU(h(s)).\label{B18}
\ee 
Thus each horizontal lift $\cC$ of $C$ is fully determined by the choice of 
an initial point $p(s_1)\in \pi^{-1}(m(s_1))$ at $s=s_1$, say,i.e., the choice 
of an element $h(s_1)\in H$. The solution to $(\ref{B18})$ is: 
\be 
\cU(h(s))=\rP \left(
\exp(-i\int_{s_1}^{s_2}\;ds^\prime\;A_{\mu}^a(q(s^\prime))\frac{dq^\mu
(s^\prime)}{ds^\prime}\; J_a)\right)\cU(h(s_1)), \label{B19}
\ee 
where $\rP$ is the path-ordering symbol keeping variables with later parameter 
values to the left. Keeping $C\subset M$ fixed and applying 
$\psi_{h_0}, \;h_0\in H$ to a horizontal lift $\cC$ of $C$ leads to another 
horizontal lift $\psi_{h_0}[\cC]$ in which $h(s)\rightarrow h(s)h_{0}^{-1}$ 
pointwise. This just amounts to changing the initial point $h(s_1)$ to 
$h(s_1)h_{0}^{-1}$.

In case $C\subset M$ is a closed loop with $q^\mu (s_2)=q^\mu (s_1)$ for a 
final parameter value $s_2$, the lift is in general not closed: while the 
end points of $\cC$ lie on the same fibre, they differ by a left translation 
by an element of $H$ determined by the loop $C$, 
\bea 
h(s_2)&=&h[C]h(s_1), \nonumber\\
\cU(h[C])=&&\rP\left(
\exp(-i\oint_{C}\;ds\;A_{\mu}^a(q(s))\frac{dq^\mu (s)}{ds} \;
J_a)\right).\label{B20} 
\eea 
These elements of the structure group $H$ form the holonomy group, in general 
a subgroup of $H$, associated with the connection $\omega$.

\vskip0.3cm \noindent {\underline{Associated Bundles}\\

To pass from the PFB $(P,M,\pi,H)$ to an associated bundle (AB) we retain the 
base space $M$, replace the typical fibre $H$ by a differentiable manifold $F$ 
as the new typical fibre, and simultaneously change the total space from $P$ 
to $E$. Thus the AB is written as a quartet $(E,M,\pi_E,F)$, with $\pi_E$ a 
new projection map $E\rightarrow M$. However it retains the memory of the 
structure group $H$ since we require that there be an action of $H$ on $F$ 
by a family of diffeomorphisms $\{\varphi_h\}$ respecting the composition 
law in $H$. Locally, $E$ looks like the Cartesian product $M\times F$, but 
this may not be so globally. We 'use up' the action of $H$ on $F$ in stating 
the transition rule connecting two overlapping local trivializations of $E$, 
in the spirit of eq.$(\ref{B2})$: 
\bea 
e\in E,&&\;\pi_E(e)=m \in
M_\alpha\subset M: e=(m,f)_\alpha,\;f\in F; \nonumber \\
&& m\in M_\alpha \cap M_\beta : e=(m,f)_\alpha =
(m,f^\prime)_\beta,\nonumber\\
&& \;\;\;\;\;f^\prime
=\varphi_{t_{\beta\alpha}(m)}(f).\label{B21}
\eea 
The transition group elements $t_{\beta\alpha}(m)$ are taken from the parent 
PFB (so implicitly the same open sets $M_\alpha, M_\beta,\cdots $ in $M$ are 
used to locally trivialize the PFB and AB). There is now no 'other' global 
fibre preserving $H$ action on $E$, in place of $\psi_h$ in the PFB; and again 
no $H$ action on $M$ is contemplated.

Compared to a general fibre bundle (FB), which however we have not recalled 
here, an AB has more structure: there is the group $H$ acting on the typical 
fibre $F$, and transition formulae connecting overlapping local 
trivializations.

If now a connection $\omega$ is given on the parent PFB, it can be used to 
set up horizontal lifts in the AB. We limit ourselves to a local coordinate 
description. With coordinates $q^\mu$ for $M$ and $f^r$ for $F$, we have 
coordinates $(q^\mu,f^r)$ for $E$. Then, given the curve $C=\{q^\mu(s)\}
\subset M$, a horizontal lift of it to E is a curve $\cC_E =\{q^\mu(s),
f^r(s)\}\subset E$ in which the coordinates $f^r(s)$ of points in the new 
fibre may be read off from eq.$(\ref{B19})$ (after taking $h(s_1)$ to be 
the identity in $H$): 
\bea 
f(s)&=& \varphi_{h(s)}(f(s_1)),\nonumber\\
\cU(h(s))&=&\rP\left( \exp(-i\int_{s_1}^{s}\;ds^\prime\;
  A_{\mu}^a(q(s^\prime))
\frac{dq^\mu
(s^\prime)}{ds^\prime}\; J_a)\right)\; 1\!\!1 .
\label{B22}
\eea

In case $F$ in the AB is a linear vector space with vectors $\Psi, 
\Psi^\prime, \cdots $,and the diffeomorphisms $\varphi_h$ are linear 
transformations on $F$ with generators $J_a$, the AB is a vector bundle; 
and horizontal lifting is describable by a simple ordinary first order matrix 
differential equation: 
\bea 
(\frac{d}{ds}&+&i A_{\mu}^a(q(s))\frac{dq^\mu (s)}{ds}\;
J_a)\Psi(s)=0, \nonumber\\
\Psi( s+\delta s) &=&\exp\left(-i A_{\mu}^a(q(s))J_a \delta
q^\mu\right)\Psi(s),\nonumber\\
\delta q^\mu &=&\frac{dq^\mu (s)}{ds}\delta s. \label{B23}
\eea
We see that both the structure group $H$ of a PFB and a connection $\omega$ 
given on it carry over to an AB set up as described above.

\vskip0.3cm \noindent {\underline{Lie group on coset space as PFB}\\

Next we consider the coset space $G/H$ of a Lie group $G$ with respect to a 
Lie subgroup $H$. We take $G/H$ to be the space of right cosets, and view $G$ 
as a PFB over it as base. We wish to point out the particular new features 
that are present in this case. For the quartet $(P,M,\pi,H)$ of a general PFB, 
we now make the identifications $P\rightarrow$ Lie group $G$, $M \rightarrow$ 
Coset space $G/H$, structure group and typical fibre $H\rightarrow$ subgroup 
$H$ in $G$. The projection $\pi$ maps any $g\in G$ to its right coset $gH$. 
So we denote such a PFB by $(G,G/H,\pi,H)$. Several new features are 
immediately recognized: the total space is a Lie group $G$, actions of $G$ on 
itself by left and right translations, $L_{g}^{(0)}$ and $R_{g}^{(0)}$, are 
both available; the former translations $L_{g}^{(0)}$ descend to a transitive 
$G$ action on the base $G/H$ by maps $L_{g}$. For the globally well defined 
fibre preserving action of $H$ on $G$ by maps $\psi_h$ we take $R_{h}^{(0)}$, 
a subset of $R_{g}^{(0)}$; so $R_{g}^{(0)}$ for $g\notin H$ play no immediate 
role.

Local trivilializations and transition formulae are connected now to choices 
of local coset representatives. Over some $M_\alpha \subset M$ a local coset 
representative is a map 
\be 
m\in M_\alpha
\rightarrow \ell_\alpha(m) \in G : \pi (\ell_\alpha(m))=m.
\label{B24} 
\ee 
This determines a local trivialization of $G$ over  $\pi^{-1}(M_\alpha)$: 
\be 
\pi(g)=m\in M_\alpha \Rightarrow g=
 \ell_\alpha(m)h, h\in H, \label{B25}
\ee
 so we can view $g$ as the pair $(m,h)_\alpha$:
 \be g=\ell_\alpha(m)h \Leftrightarrow
 g=(m,h)_\alpha .\label{B26}
 \ee
In the overlap of two such choices of local coset representatives we 
necessarily have
 \be 
 m\in M_\alpha\cap M_\beta :\ell_\beta(m)=\ell_\alpha(m)
 t_{\alpha\beta}(m),\;\;t_{\alpha\beta}(m)\in H, \label{B27}
\ee
 so that 
\be 
g=(m,h)_\alpha=(m^\prime,h^\prime)_\beta
 \Leftrightarrow h^\prime =t_{\beta\alpha}(m)h. \label{B28}
 \ee 
In this kind of PFB, there is a preferred connection. We have denoted a basis 
for ${\underline H}$ by $\{e_a\}$. They obey Lie bracket relations
\be 
[e_a,e_b]={C_{ab}}^c e_c, \label{B29}
\ee 
with ${C_{ab}}^c$ the structure constants of $H$. Now we add elements $e_\mu$ 
to get a basis for ${\underline G}$
\be 
G= {\rm Sp}\{e_a,e_\mu\}.\label{B30}
\ee
 We assume that the Lie brackets $[e_a,e_\mu]$ have the simple form
\be 
[e_a,e_\mu]={C_{a\mu}}^\nu e_\nu, \label{B31}
\ee
with no $e_b$ terms on the right. The remaining Lie bracket
relations for ${\underline G}$ are
 \be 
[e_\mu,e_\nu]={C_{\mu\nu}}^a e_a+{C_{\mu\nu}}^{\lambda} e_\lambda . \label{B32}
\ee
The full set of left-invariant Maurer-Cartan one-forms on $G$ consists of 
$\hat{\Theta}^{(0)a}(g), \hat{\Theta}^{(0)\mu}(g)$. It now turns out that 
a preferred connection on $(G,G/H,\pi,H)$ is given by 
\be 
\omega=\hat{\Theta}^{(0)a}(g)e_a , \label{B33}
\ee 
which is indeed ${\underline H}$-valued. Here only a subset of the full set of 
Maurer-Cartan forms on $G$ is used. This choice obeys both the conditions 
demanded of a general connection on a PFB. If now one brings in a local coset 
representative $\ell_\alpha(m)$ and the accompanying local trivialization 
$(\ref{B26})$ of $G$, one indeed finds 
\bea 
&&g=\ell_\alpha(m)h=(m,h)_\alpha :\nonumber\\
&&\hat{\Theta}^{(0)a}(g)
=\hat{\Theta}_{H}^{(0)a}(h)-{{{\cD}_H}^{a}}_b(h^{-1})A^b(m),
\label{B34}\eea 
where $A^a$ are specific one-forms over $M_\alpha$ arising out of the
structure constants ${C_{a\mu}}^\nu, {C_{\mu\nu}}^a,
{C_{\mu\nu}}^\lambda$. Here the Maurer-Cartan forms and adjoint representation
matrices belonging to $H$ have been denoted with a subscript $H$ to
distinguish them from objects belonging to $G$. Using $(\ref{B34})$ in
$(\ref{B33})$ we see that the expected structure 
$(\ref{B7})$ for $\omega$ is indeed present, so this is a preferred connection 
determined by $G$ in relation to $G/H$ and $H$.

\vskip0.3cm \noindent {\underline{Associate bundle to a coset PFB}\\

Lastly we point out the special features that accompany a bundle 
$(E,G/H,\pi_E,F)$ associated to a coset space PFB $(G,G/H,\pi,H)$. 
They share the same base $G/H$, while the total space and the typical fibre 
are $E$ and $F$ in place of $G$ and $H$ respectively. As with a general AB, 
we have $H$ acting on $F$ via diffeomorphisms $\varphi_h$. The transition 
functions $\{t_{\alpha\beta}(m)\}$ belonging to $(G,G/H,\pi,H)$ are used for 
$(E,G/H,\pi_E,F)$ as well. Compared to a general AB in which eq. $(\ref{B21})$ 
holds, we now have the feature that the transition functions arise from coset 
representatives via eq.$(\ref{B27})$, namely 
\be 
t_{\alpha\beta}(m)= \ell_\alpha(m)^{-1}\ell_\beta(m), \label{B35}
\ee 
though the $\ell_\alpha(m)$ themselves play no direct role in the AB. The 
left and right translations $L_{g}^{(0)}, R_{g}^{(0)}$ of $G$ also have no 
role to play, though the present AB `remembers' (if necessary) the transitive 
$G$ action on the base $G/H$ by maps $L_g$. Finally the preferred connection 
on $(G,G/H,\pi,H)$ given by eq.$(\ref{B33})$ leads to parallel transport and 
horizontal lifting operations in $(E,G/H,\pi_E,F)$ in a preferred manner, the 
details being of course given by eqs.$(\ref{B22},\ref{B23})$.


\begin{thebibliography}{50}
\bibitem{1} M. V. Berry, Proc. Roy. Soc. A{\bf 392}, 45 (1984).
 Many of the early papers on geometric phase have been 
reprinted in  {\it Geometric Phases in 
Physics} by A. Shapere and F. Wilczek,, (World Scientific, Singapore, 
1989) and in {\it Fundamentals of Quantum Optics}, SPIE Milestone 
Series, edited by G. S. Agarwal, (SPIE Press, Bellington, 1995).
\bibitem{2} Y. Aharanov and J. Anandan, Phys. Rev. Lett. {\bf 58}, 1593 (1987).
\bibitem{3} J. Samuel and R. Bhandari, Phys. Rev. Lett. {\bf 60}, 
2339 (1988).
\bibitem{4} N. Mukunda and R. Simon Ann. Phys. (NY) {\bf 228}, 205 
(1993); {\it ibid} {\bf 228}, 269 (1993).
\bibitem{5} B. Simon Phys. Rev. Lett. {\bf 51}, 
2167 (1983).
\bibitem{6} A. Uhlmann Rep. Math. Phys. {\bf 24}, 229 (1986);
{\it ibid}, {\bf 36}, 461 (1995). 
\bibitem{7} E. Sj\"oqvist, A. K. Pati, A. Ekert, J. S. Anandan, M. 
Ericsson, D. K. Loi and V. Vedral, Phys. Rev. Lett. {\bf 85}, 
2845 (2000).
\bibitem{8} L. Dabrowski and A. Jadczyk, J. Phys A{\bf 22} , 3167 (1989). 
\bibitem{9} A. A. Kirillov, Bull. Am. Math. Soc. {\bf 36}, 433 (1999).
\bibitem{10} E. Ercolessi, G. Marmo, G. Morandi and N. Mukunda, Int.J. 
Mod. Phys. A{\bf 16}, 5007 (2001). 
\bibitem{11} A. P. Balachandran, G. Marmo, B.-S. Skagerstam and A. 
Stern, {\it Gauge Symmetry and Fibre Bundles- Applications to 
Particle Dynamics} (Springer, Berlin, 1983). 
\bibitem{12}A. P. Balachandran, G. Marmo, B.-S. Skagerstam and A. 
Stern, {\it Classical Topology and Quantum States}, 
 (World Scientific, Singapore, 1991). 
\bibitem{13} E. M. Rabei, Arvind, N. Mukunda, and R. Simon, Phys. Rev. 
A{\bf 60}, 3397 (1999). 
\bibitem{14}  N. Mukunda, Arvind, E. Ercolessi, G. Marmo, G. Morandi and 
R. Simon, Phys. Rev. 
A{\bf 67}, 042114 (2003). 
\bibitem{15} V. I. Arnold, {\it Mathematical Methods of Classical 
Mechanics}, (Springer, Berlin, 1978), Appendices 2 and 5. 
\bibitem{16} M. A. Nielsen Phys. Rev. A{\bf 62}, 052308 (2000). 
\bibitem{17} F. Pistolesi and N. Manini, Phys. Rev. Lett. {\bf 85}, 
1585 (2000); {\bf 85}, 3067 (2000). 
\bibitem{18} N. Mukunda, Arvind, S. Chaturvedi and R. Simon, Phys. Rev. 
A{\bf 65}, 012102 (2001). 
\bibitem{19} F. Wilczek and A. Zee, Phys. Rev. Lett. {\bf 52}, 2111 
(1984).
\bibitem{20} Y. Choquet-Bruhat and  C. Dewitt-Morette and M.  Dillard-Bleick,
{\it Analysis, Manifolds and Physics-Part 1:Basics}, revised edition, (North
Holland, Amsterdam 1991).  
\bibitem{21} S. Kobayashi and K. Nomizu, {\it Foundations of 
Differential Geometry}, ( Interscience, New York, 1969). 
\bibitem{22} C. Nash and S. Sen, {\it Topology and Geometry for 
Physicists}, (Academic Press, 1983). 
\bibitem{23} N. Mukunda, {\it Geometrical Methods for Physics} in 
{\it Geometry, Fields and Cosmology} eds. B. R. Iyer and C. V. 
Vishveshwara, (Kluwer, Dordrecht, 1997).
\end{thebibliography}
\end{document}